\documentclass[epj]{svjour}
\usepackage{graphics}
\usepackage{graphicx}
\usepackage[T1]{fontenc}
\usepackage[latin1]{inputenc}
\begin{document}
\title{Spherical Hartree-Fock calculations with linear 
momentum projection before the variation.}
\subtitle{Part I: Energies, form factors , charge densities and 
mathematical  sum rules.}
\author{R.R. Rodr\'{\i}guez--Guzm\'an and K.W. Schmid
}                     
\mail{rayner@tphys.physik.uni-tuebingen.de}
\institute{Institut für Theoretische Physik der Universität Tübingen, Auf der
Morgenstelle 14, D-72076 Tübingen, Germany.}
\date{Received: date / Revised version: date}
% The correct dates will be entered by Springer
%
\abstract{Spherical Hartree--Fock calculations with projection onto zero total linear
momentum before the variation are performed for the nuclei $^4$He, $^{12}$C,
$^{16}$O, $^{28}$Si, $^{32}$S and $^{40}$Ca using a density--independent
effective nucleon--nucleon interaction. The results are compared to those of
usual spherical Hartree--Fock calculations subtracting the kinetic energy of
the center of mass motion either before or after the variation and to the
results obtained analytically with oscillator occupations. Total energies,
hole--energies, elastic charge form factors and charge densities and the
mathematical Coulomb sum rules are discussed.
} %end of abstract
\authorrunning{R.R. Rodr\'{\i}guez--Guzm\'an and K.W. Schmid}
\titlerunning{Spherical Hartree-Fock calculations with linear  momentum ...}
\maketitle

\section{Introduction}
We consider the nucleus as a closed system of interacting, non--relativistic
nucleons. The homogenity of space requires that the total linear momentum of
this system is conserved. Consequently the hamiltonian describing any
particular nucleus cannot depend on the center of mass (COM) coordinate of its
constituents, but (besides on spin-- and isospin--quantum numbers) only
on relative coordinates and momenta. The dependence on the total momentum
is trivial : it describes the free motion of the total system and can
always be transformed away by considering the system in its COM rest frame.
We have then to solve the corresponding Schr\"odinger equation for the
remaining ``internal'' hamiltonian. In principle this can be achieved by
writing this hamiltonian in Jacobi coordinates. However, nucleons are fermions
and thus do obey the Pauli principle. Since the Jacobi coordinates depend on
all the nucleon coordinates, thus an explicit antisymmetrization of the wave
functions is required as it is performed, e.g., in few--body physics. Being
already there sometimes rather involved though still feasible, such an
explicit antisymmetrization becomes impossible in the many--body system 
(e.g., the antisymmetrization of 20 like nucleons would require 20 factorial
different terms). Thus in the many--body system the antisymmetrization usually
is performed implicitely by expanding the wave functions in terms of Slater
(or generalized Slater) determinants. In this way the Pauli principle is
automatically fulfilled. The Slater determinants, however, depend on 3A
instead of the allowed 3A-3 coordinates and thus contain contaminations due to
the motion of the system as a whole, so called ``spurious'' admixtures.
Galilei--invariance is broken.

This defect of almost all microscopic nuclear structure models has been
recognised \cite{ref1.} almost immediately after the development of the shell--model.
It was shown later on \cite{ref2.} that in case of pure harmonic oscillator
configurations one can get rid of this problem by diagonalizing the
(oscillator) COM hamiltonian and projecting all states not corresponding
to the ground state of this operator out of the spectrum of the many nucleon
hamiltonian. This procedure, however, requires the use of so called complete
$n\hbar\omega$--spaces (since only then COM and internal excitations 
decouple exactly) and thus is of little help in most of the usual approaches
to the nuclear many--body problem. A more general solution is the projection
of the wave functions into the COM rest frame \cite{ref3.}, which ensures
translational-- and, if performed before solving the corresponding
Schr\"odinger equation (usually by variational methods) even full
Galilei--invariance \cite{ref4.}. The key idea of this projection is to superpose
the wave function shifted all over normal space with identical weights
and thus to achieve vanishing total linear momentum. Since the bound states
of a nucleus are localized, this procedure always does converge (for
scattering states a slightly different procedure has to be used \cite{ref13.}).
The projection method has the advantage that it works in general model
spaces as well as for general (non--oscillator) wave functions.

\begin{figure*}
\begin{center}
\includegraphics[angle=-90,width=12cm]{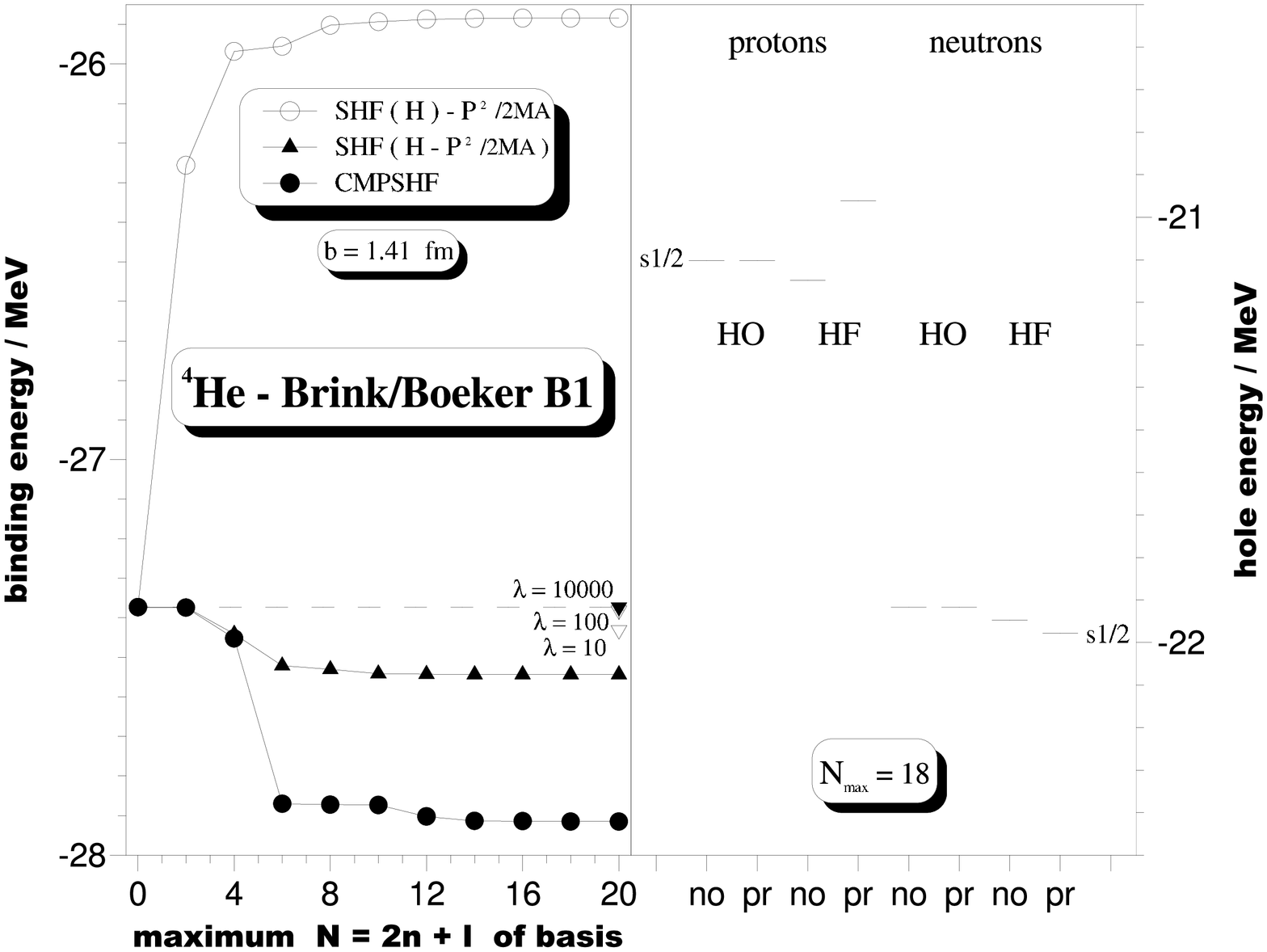}
\caption{In the left part the total binding energy of $^4$He as obtained
with the Brink--Boeker force B1 [14] using an oscillator length b=1.41 Fm
is plotted against the size of the basis. Three curves are shown : the
open circles correspond to a normal spherical Hartree--Fock calculation
with the expectation value of $P^2/2MA$ subtracted after convergence,
the full triangles display the results if this center of mass correction
is included during the iteration, and, finally, the full circles give
the result of a spherical Hartree--Fock calculation with projection into
the momentum rest frame before the variation (CMPSHF). Furthermore, the
figure displays the results of constrained calculations adding the
oscillator center of mass Hamitonian with a Lagrangian multiplier
$\lambda$ to the internal hamiltonian which ``penalizes'' COM excitations.
The right part of the figure displays the hole energies obtained in the
oscillator limit HO (N$_{\rm max}$=0) and in the Hartree--Fock approximation HF
(with N$_{\rm max}$=18) using the ``normal'' (no) as well as the ``projected'' (pr) approach in both cases. Note that the normal approach includes the usual
COM correction (see text).}
\end{center} 
\end{figure*}

Though in principle known since almost half a century, only few practical
calculations have been performed using this method. The reason for this
is quite simple : the projection operator is an A--body integral operator with
the rather nasty property to link the usual model space states to rather
highly excited (and thus usually unoccupied) ones as well as to the fully
occupied ones, which are often treated as an inert core. This is easy
to understand : any change of the linear momentum of the valence nucleons
requires a corresponding change of the linear momentum of the core in order
to ensure vanishing total linear momentum for the system. Unlike angular
momentum, linear momentum is thus a true A--body correlation and hence
much more complicated to treat than the latter. 

\begin{figure*}
\begin{center}
\includegraphics[angle=-90,width=12cm]{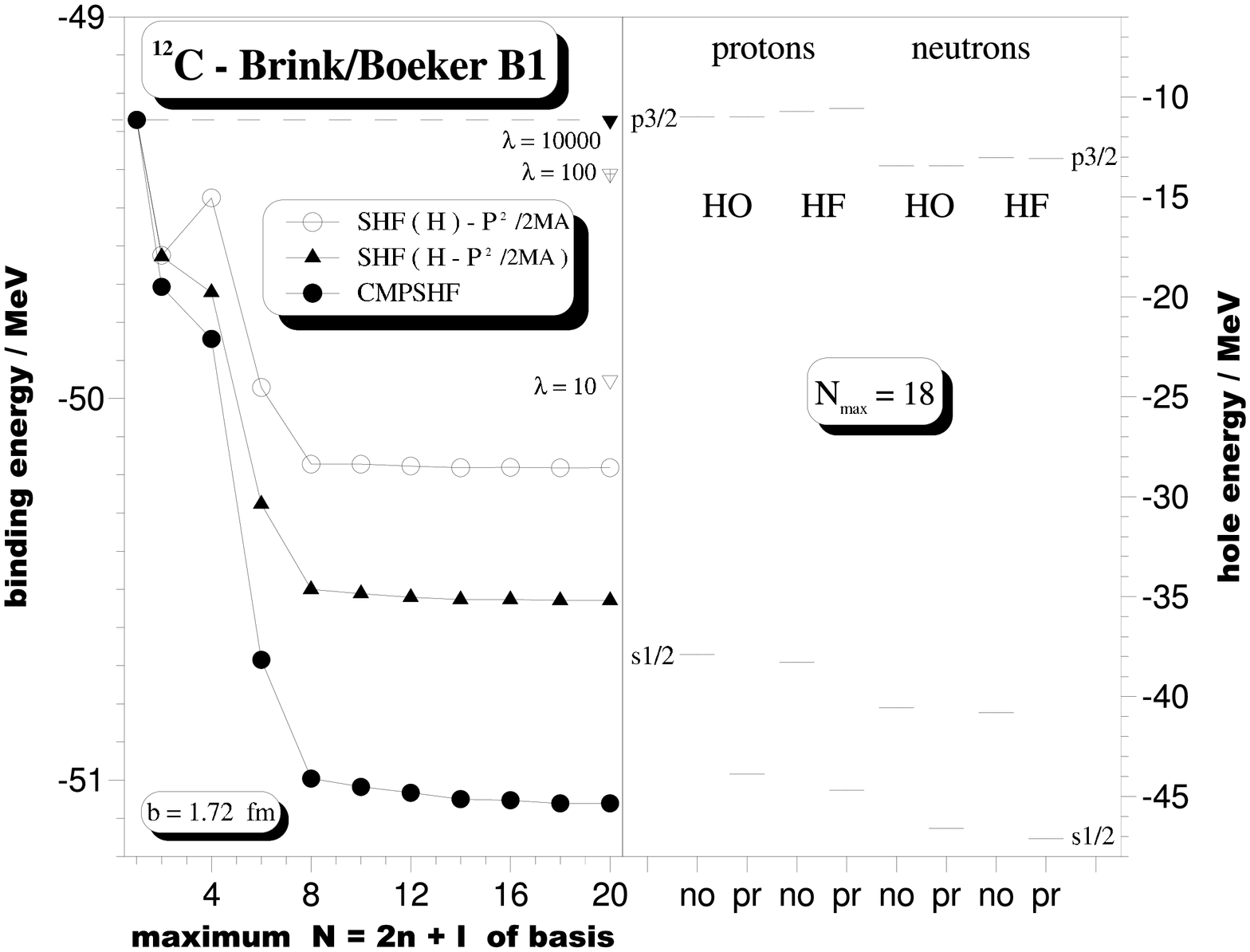}
\caption{Same as in Fig. 1, but for the nucleus $^{12}$C with
oscillator length b=1.72 Fm.}
\end{center} 
\end{figure*}

\begin{figure*}
\begin{center}
\includegraphics[angle=-90,width=12cm]{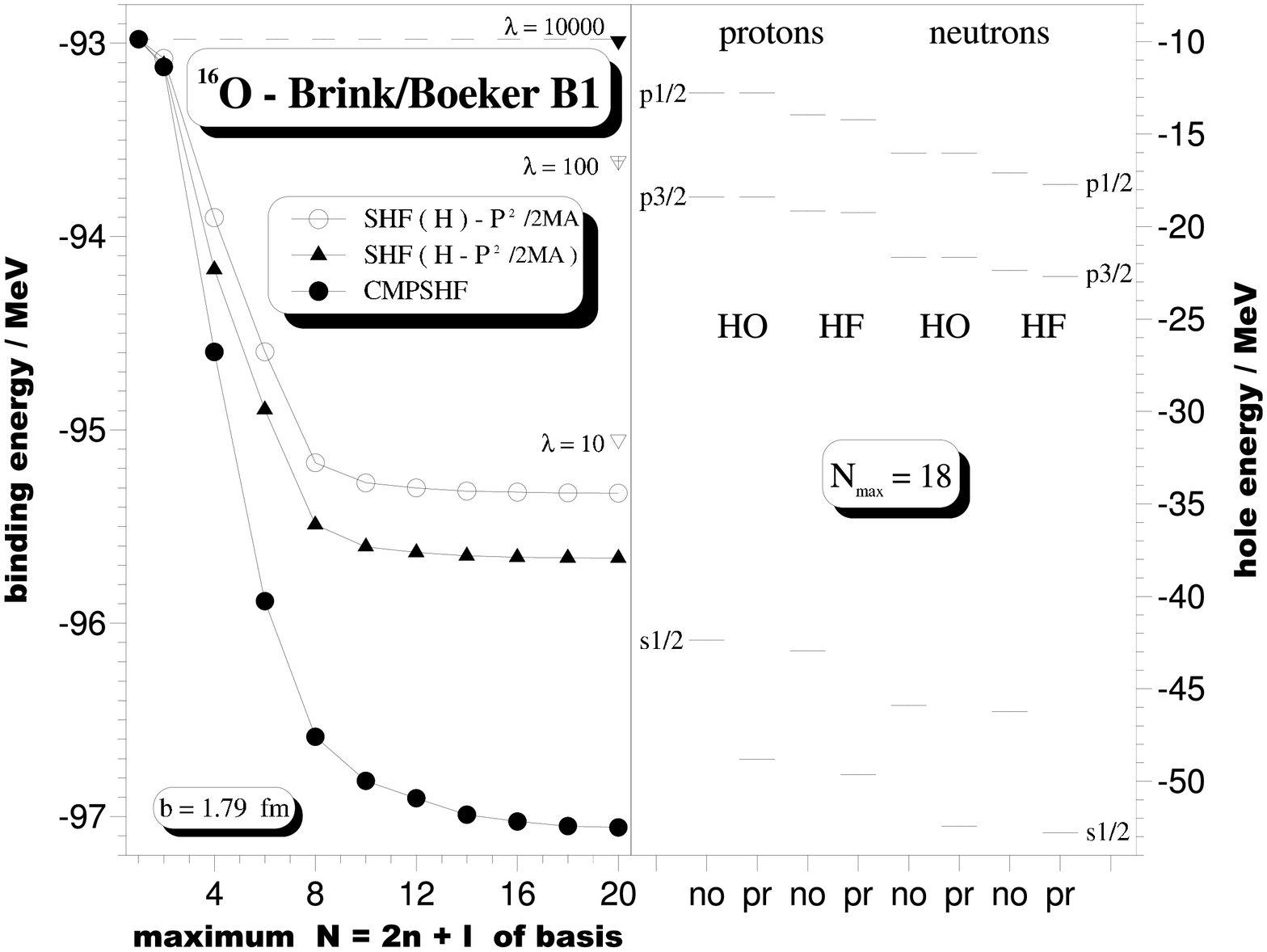}
\caption{Same as in Fig. 1, but for the nucleus $^{16}$O with
oscillator length b=1.79 Fm.}
\end{center} 
\end{figure*}

Because of these difficulties, instead of treating
Galilei--invariance correctly, its breaking is usually neglected
adopting the well known text book argument that it induces only 1/A
effects and thus ``can savely be neglected for nuclei beyond oxygen'' 
\cite{ref4.}, provided the usual approximate corrections like subtracting 
the kinetic
energy of the COM motion from the original hamiltonian or the use of the
so called Tassie--Barker factor \cite{ref5.} in the analysis of form factors are
done. 

\begin{figure*}
\begin{center}
\includegraphics[angle=-90,width=12cm]{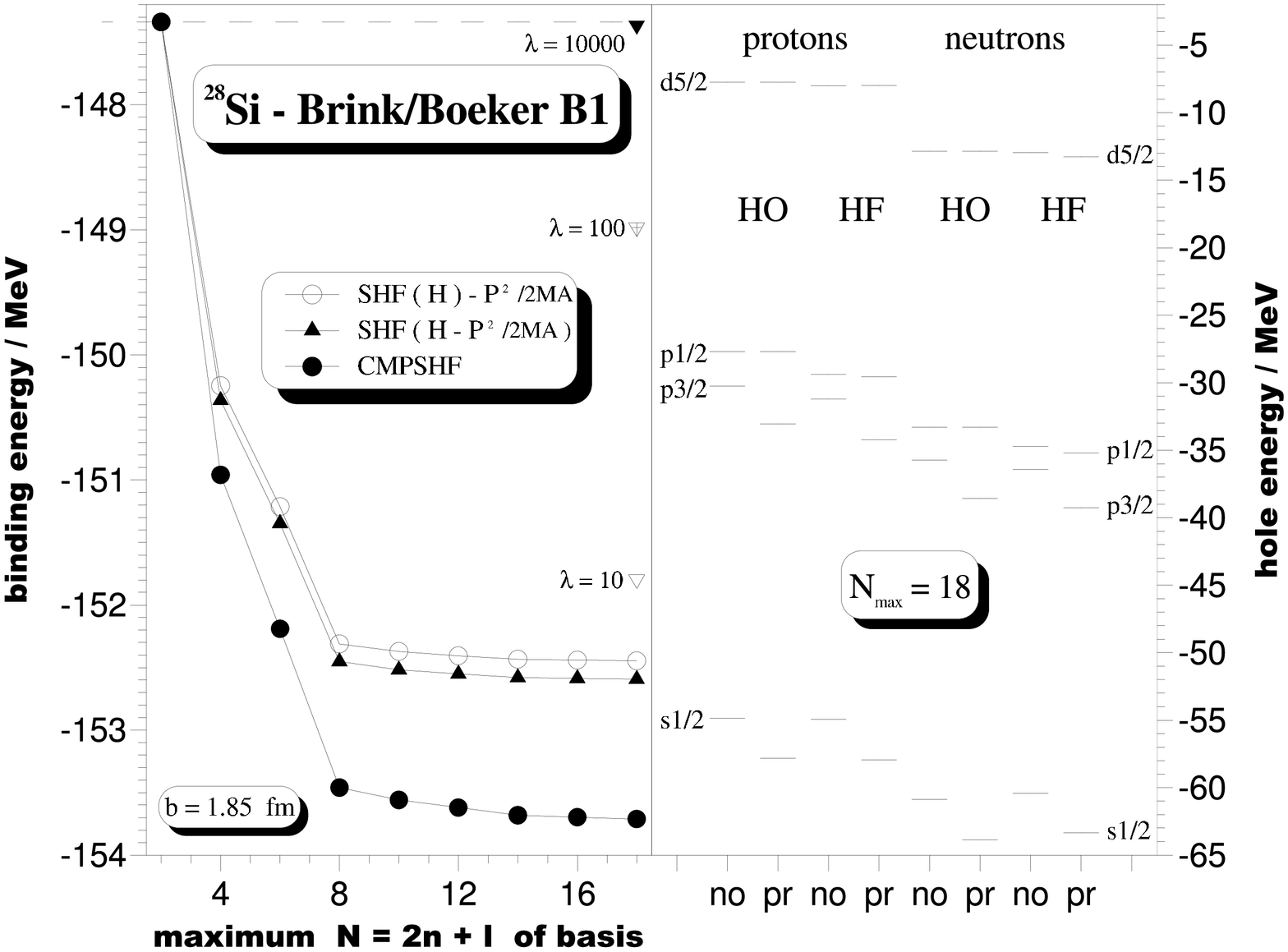}
\caption{Same as in Fig. 1, but for the nucleus $^{28}$Si with
oscillator length b=1.85 Fm.}
\end{center} 
\end{figure*}

\begin{figure*}
\begin{center}
\includegraphics[angle=-90,width=12cm]{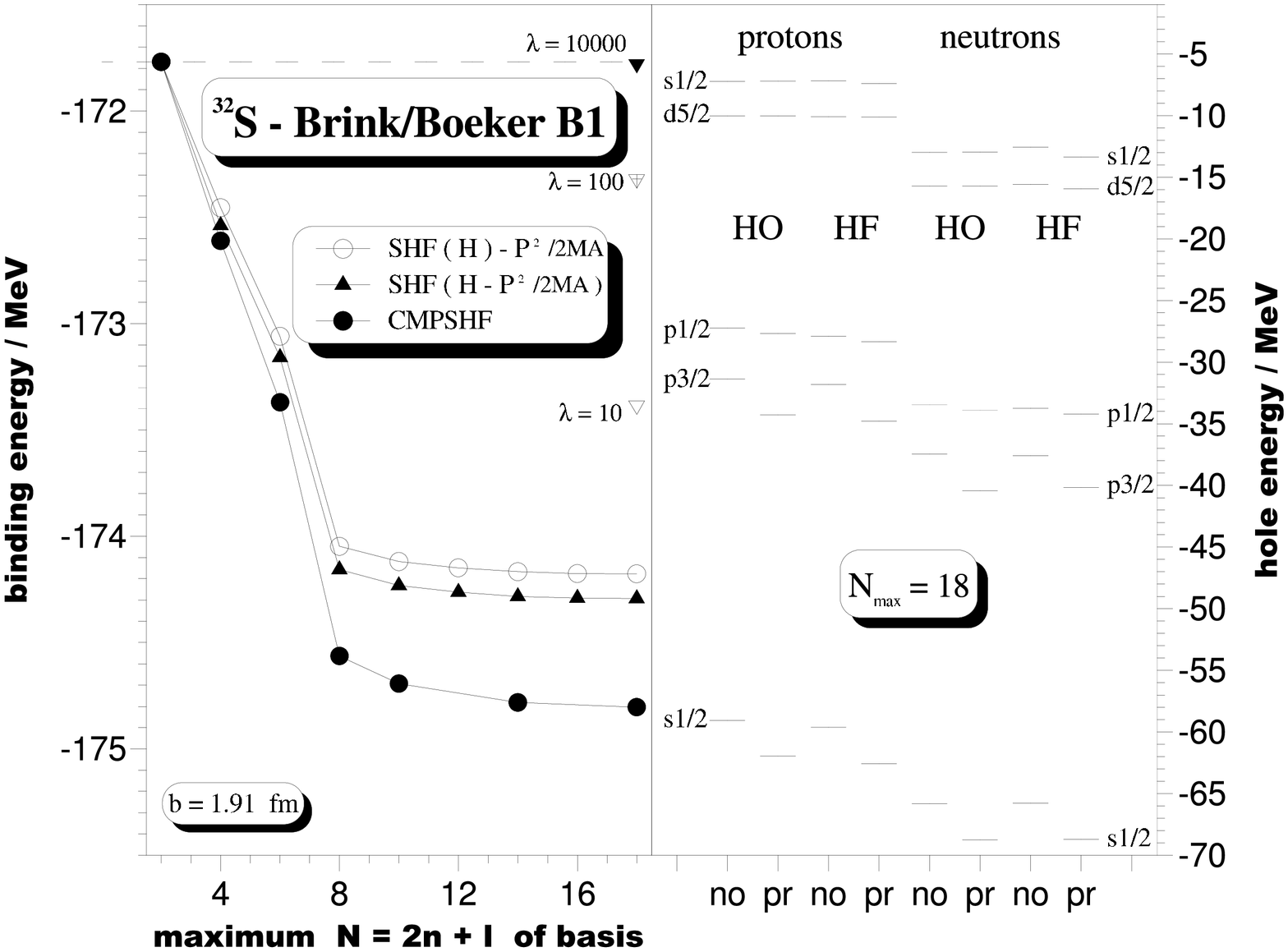}
\caption{Same as in Fig. 1, but for the nucleus $^{32}$S with
oscillator length b=1.91 Fm.}
\end{center} 
\end{figure*}

\begin{figure*}
\begin{center}
\includegraphics[angle=-90,width=12cm]{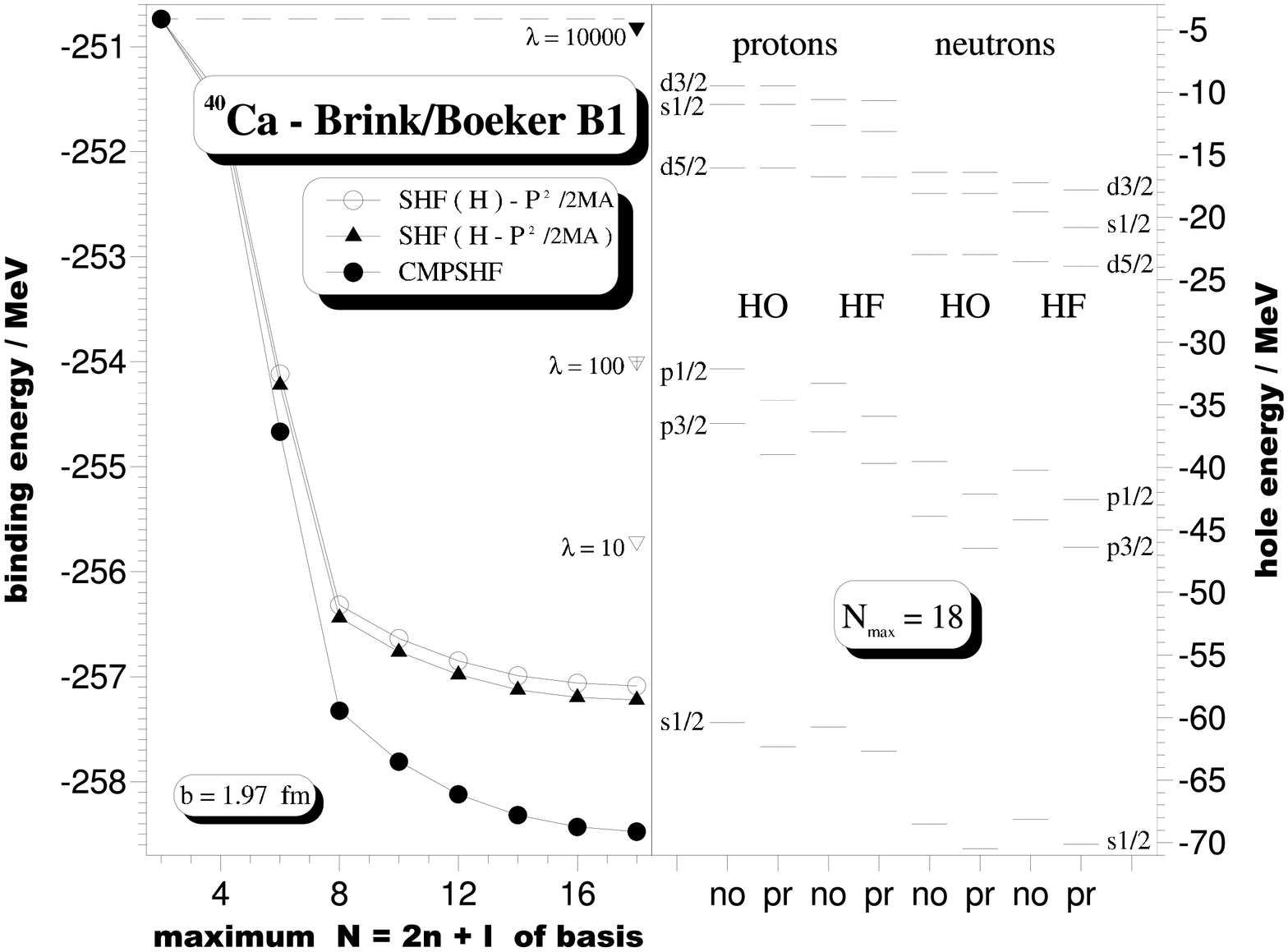}
\caption{Same as in Fig. 1, but for the nucleus $^{40}$Ca with
oscillator length b=1.97 Fm.}
\end{center} 
\end{figure*}

That this, however, is not true has been shown by several studies within
the last decade. Hartree--Fock calculations with projection into
the COM rest frame for $^4$He [6] as well as the analysis of form factors
and charge densities of several spherical nuclei \cite{ref7.,ref8.}
 have demonstrated
that the correct treatment of Galilei--invariance yields considerable
effects far beyond the usually assumed 1/A level. The same holds
 for scattering
states as demonstrated in ref. \cite{ref9.} for the inclusive 
quasi--elastic
electron scattering again from $^4$He. Recently now, a whole series
of model investigations \cite{ref10.,ref11.,ref12.,ref13.} has been published, in which the COM
effects have been studied in a more systematic way. Considerable effects
have been seen for spectral functions and spectroscopic factors,
transition form factors and densities, energies of hole--states,
Coulomb sum rules, response functions and many more. These investigations,
however, have been undertaken with rather simple wave functions :
the ground states of the doubly even A--nucleon systems $^4$He, $^{16}$O
and $^{40}$Ca have been described in the simple oscillator limit and
for the ground and excited states of the corresponding odd (A-1)--nucleon
systems simple one--hole states have been used. This has the advantage
that all calculations can be performed analytically but is definitely
not very realistic. So, e.g., the above mentioned pure oscillator A--nucleon
configurations are non--spurious and thus the projection yields here no
additional effect with respect to the usual approach to subtract the kinetic
energy of the COM motion. It is hence desirable to study, e.g., these
ground states in more realistic approaches. This will be done in this and
a forthcoming paper.

For this purpose we have performed spherical Hartree--Fock calculations
with projection
into the COM rest frame before the variation for the six nuclei $^4$He,
$^{12}$C, $^{16}$O, $^{28}$Si, $^{32}$S and $^{40}$Ca. The results have
been compared with those of normal spherical Hartree--Fock calculations
subtracting the kinetic energy of the COM motion either before or after
the variation and with the analytically obtained oscillator results out of
ref. \cite{ref12.}. For each of the considered nuclei up to 19 major oscillator
shells have been used as single particle basis. As effective interaction
the simple Brink--Boeker force B1 \cite{ref14.} has been taken. We are aware of the
fact that this interaction is not very realistic. However, the aim of the
present investigation is not a comparison with experiment but the study
of the effects of a correct treatment of Galilei--invariance. For this
purpose the B1--interaction is as good as any other. Furthermore, consisting
out of Gaussians, it can be treated in the oscillator limit analytically
and thus allows for a direct comparison with the results reported in ref.
\cite{ref12.}.

Section 2 of the present paper gives a short summary of the spherical
Hartree--Fock approach with projection into the COM rest frame before
the variation. Section 3 will then describe some details of the calculations
and present the results for the total energies, the hole--energies, the
elastic charge form factors and corresponding charge densities and the
Coulomb sum rules. Conclusions, three appendices with some detailed
formulas and references conclude the present paper.

In the second of the present series of two papers we shall then discuss
the effects of the correct treatment of Galilei--invariance on the spectral
functions and spectroscopic factors obtained with the wave functions
out of the present paper.

\section{COM-projected Hartree-Fock.}
The essential mathematics for Hartree--Fock calculations
with projection into the COM rest frame before the variation has
been presented in detail already in ref. [6] and hence will be summerized
only briefly in the following. We start by defining our model space by
$M_b$ oscillator single particle states, the creators of which will be
denoted by $\{c^{\dag}_i\,;\;i=1,...,M_b\}$. We shall furthermore assume
that the effective hamiltonian appropriate for this model space is
known and can be written in the chosen representation as a sum of
only one-- and two--body parts

\begin{equation} \label{Eq1}
\hat H\;=\;\sum_{ir} t(ir) c^{\dag}_i c_r\;+\;
\sum_{ikrs} v(ikrs) c^{\dag}_i c^{\dag}_k c_s c_r 
\end{equation}
where $t(ir)$ are the single particle matrix elements of the kinetic
energy operator and $v(ikrs)$ the antisymmetrized two--body matrix elements
of the considered interaction. We shall assume that this interaction
is translational invariant, i.e., it does not depend on the center of mass
coordinate of the two nucleons. Density dependent interactions (in their
usual form) do not fulfill this requirement. Their treatment is much more
complicated as has been described in detail in ref. \cite{ref12.}. Such interactions
will not be considered in the present paper.

In the Hartree--Fock approach one searches for the optimal one--determinant
representation of the A--nucleon ground state having the form

\begin{equation} \label{Eq2}
\vert D\rangle\;=\;\left\{\prod_{h=1}^A\;b^{\dag}_h\;\right\}
\vert 0\rangle 
\noindent
\end{equation}
where
\begin{equation} \label{Eq3}
b^{\dag}_\beta\;=\;\sum_{i=1}^{M_b}\; D^*_{i\beta}
c^{\dag}_i
\end{equation}
and 

\begin{equation} \label{Eq4}
b_\beta\;=\;\sum_{i=1}^{M_b}\; D_{i\beta}
c_i 
\end{equation}
respectively, with $D$ being a unitary $(M_b\times M_b)$ transformation.
In eq. (2) we have assumed that the selfconsistent states created by
the operators (3) are ordered according to their energy so that $\beta=1,...,A$
correspond to the occupied ``hole'' states $h,\;h',\;...$. The
unoccupied orbits $\beta=A+1,...,M_b$ will be denoted as
``particle'' states $p,\;p',\;...$ in the following.

\begin{figure*}
\begin{center}
\includegraphics[angle=-90,width=12cm]{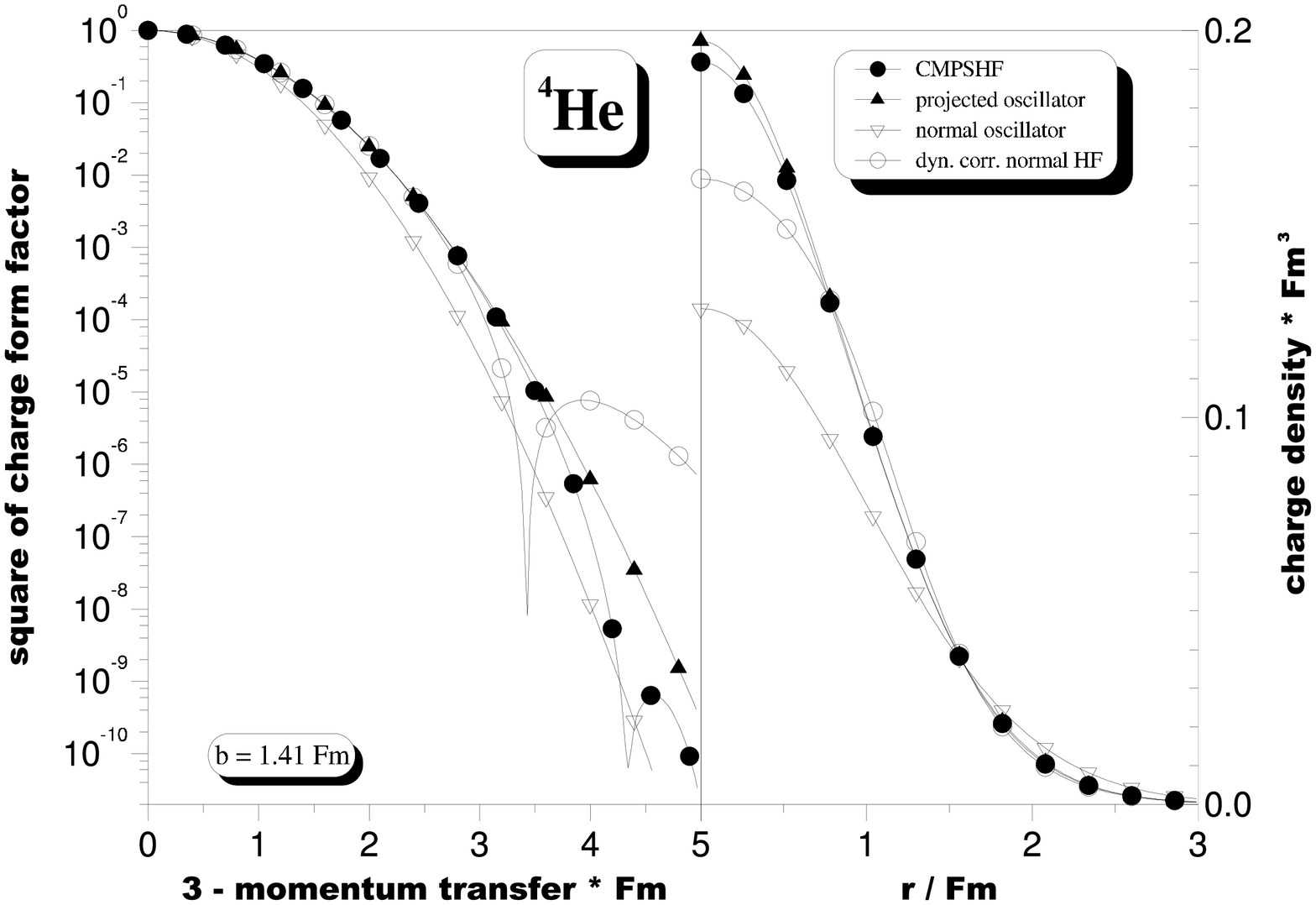}
\caption{In the left part of the figure the square of the charge
form factor for $^4$He is displayed as function of the 3--momentum
transfer. For the nucleon form factors the usual dipole parametrisation
[21] has been used. Open inverted triangles correspond to an oscillator
occupation with no COM correction included (normal oscillator), full triangles give the
oscillator result including the Tassie--Barker factor [5] (projected oscillator),
open circles display the form factor obtained with normal Hartree--Fock
(including the COM correction in the hamiltonian during the variation) taking
into account the dynamic correction (essentially again the Tassie--Barker factor)
out of the text. Finally, the full circles display the result of the projected
calculation. The right part of the figure gives the corresponding charge densities
(obviously calculated for point nucleons).}
\end{center} 
\end{figure*}

Now, obviously, the determinant (\ref{Eq2}) is not translationally invariant. In order
to obtain a Galilei--invariant wave function we have to use instead of (\ref{Eq2}) the
expression

\begin{equation} \label{Eq5}
\vert D\,;\;0\rangle\,\equiv\,{{\hat C(0)\vert D\rangle}\over
{\sqrt{\langle D\vert\hat C(0)\vert D\rangle}}}
\noindent
\end{equation}
as test wave function in the variation. Here
\begin{equation} \label{Eq6}
\hat C(0)\,\equiv\,\int d^3\,\vec a\,\hat S(\vec a\,)
\end{equation}
with
\begin{equation} \label{Eq7}
\hat S(\vec a\,)\,\equiv\,\exp\{i\vec a\cdot\hat P\}
\end{equation}
projects into the COM rest frame by superposing all states created
by the shift operator (\ref{Eq7}) (here $\hat P$ is the operator of the total
momentum of the considered system) with identical weights.

\begin{figure*}
\begin{center}
\includegraphics[angle=-90,width=12cm]{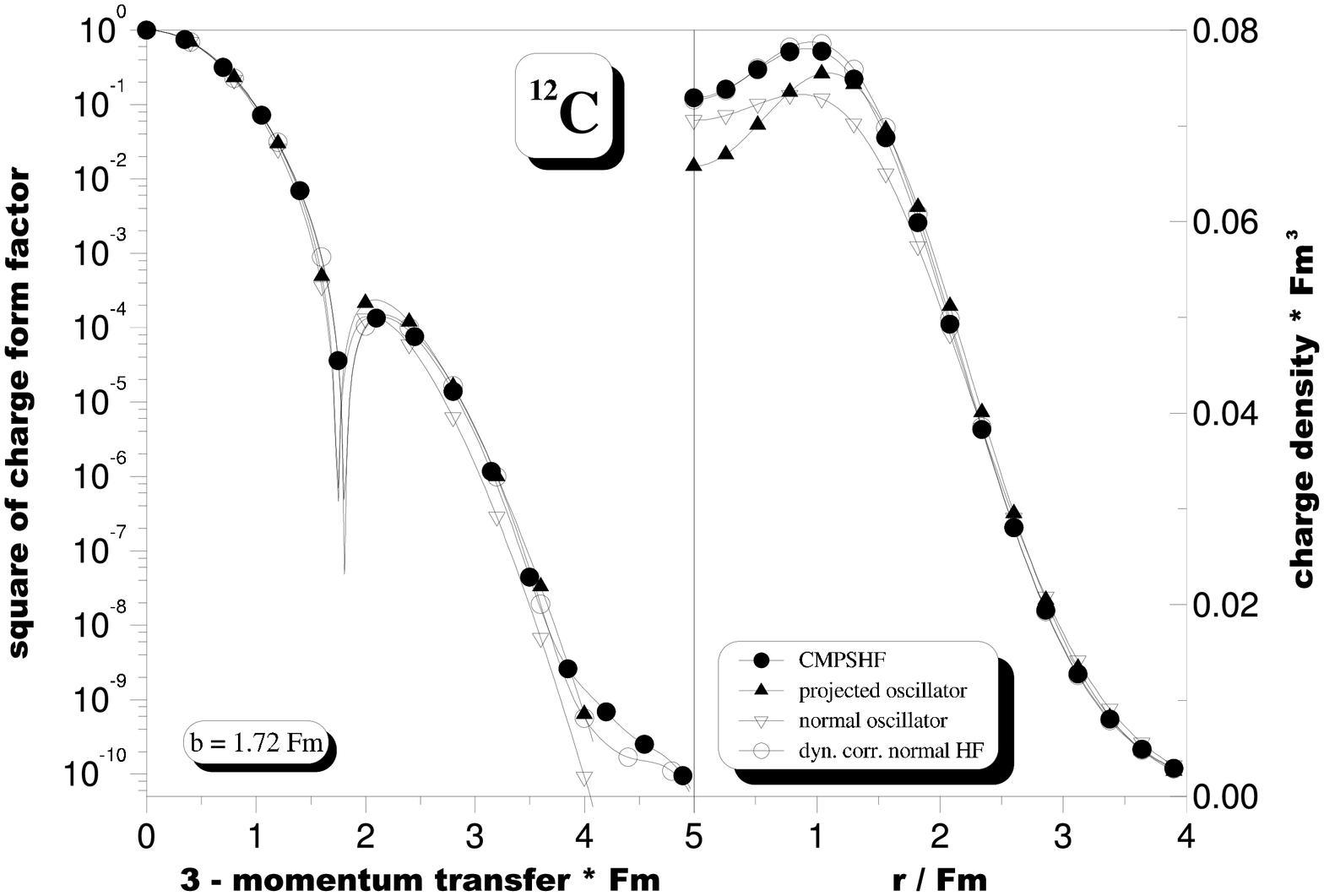}
\caption{Same as in Fig. 7, but for the nucleus $^{12}$C.}
\end{center} 
\end{figure*}

The energy functional $E_{pr}$ to be used in the
Hartree--Fock approach with projection into the COM rest frame before
the variation can then be written as
\begin{equation} \label{Eq8}
E_{pr}\;=\;{{\int d a\;a^2\int d\Omega_a\;
h(\vec a\,)}\over{
\int d a\;a^2\int d\Omega_a\;n(\vec a\,)}}
\end{equation}
where we have introduced the shifted overlap function
\begin{equation} \label{Eq9}
n(\vec a\,)\;\equiv\;\langle D\vert\hat S(\vec a\,)\vert D\rangle\;
=\;{\rm det} X(\vec a\,) 
\end{equation}
which can be represented as the
determinant of an $(A\times A)$-matrix
\begin{equation} \label{Eq10}
X_{hh'}(\vec a\,)\;\equiv\;\langle h\vert\hat S(\vec a\,)\vert
h'\rangle\;=\;\sum_{ik}\;
D_{ih} S_{ik}(\vec a\,) D^*_{kh'}
\noindent
\end{equation}
with $S_{ik}(\vec a\,)$ being the matrix representation of
the shift operator within the chosen harmonic oscillator
single particle basis. These matrix elements are
given in the appendix A. Furthermore we use in (\ref{Eq8})
the shifted energy function
\begin{equation} \label{Eq11}
h(\vec a\,)\;\equiv\;\langle D\vert\hat H\hat S(\vec a\,)\vert
D\rangle\;=\;t(\vec a\,)\;+\;v_2(\vec a\,) 
\end{equation}
Here the one--body term is given by
\begin{equation} \label{Eq12}
t(\vec a\,)\;=\;n(\vec a\,)\;
\sum_{ir}\;t(ir)\;\tilde\rho_{ri}(\vec a\,)
\noindent
\end{equation}
with the shifted density matrix being defined as
\begin{equation} \label{Eq13}
\tilde\rho_{ri}(\vec a\,)\;\equiv\;
\sum_k\;S_{rk}(\vec a\,)\;\sum_{hh'}\;
D^*_{kh}\;[X^{-1}(\vec a\,)]_{hh'}\;D_{ih'} 
\noindent
\end{equation} 
and for the two-body part of (\ref{Eq11}) one obtains
\begin{equation} \label{Eq14}
v_2(\vec a\,)\;=\;n(\vec a\,)\;{1\over 2}\sum_{ikrs}\;
v(ikrs)\;
\tilde\rho_{sk}(\vec a\,)\;\tilde\rho_{ri}(\vec a\,) 
\end{equation} 

\begin{figure*}
\begin{center}
\includegraphics[angle=-90,width=12cm]{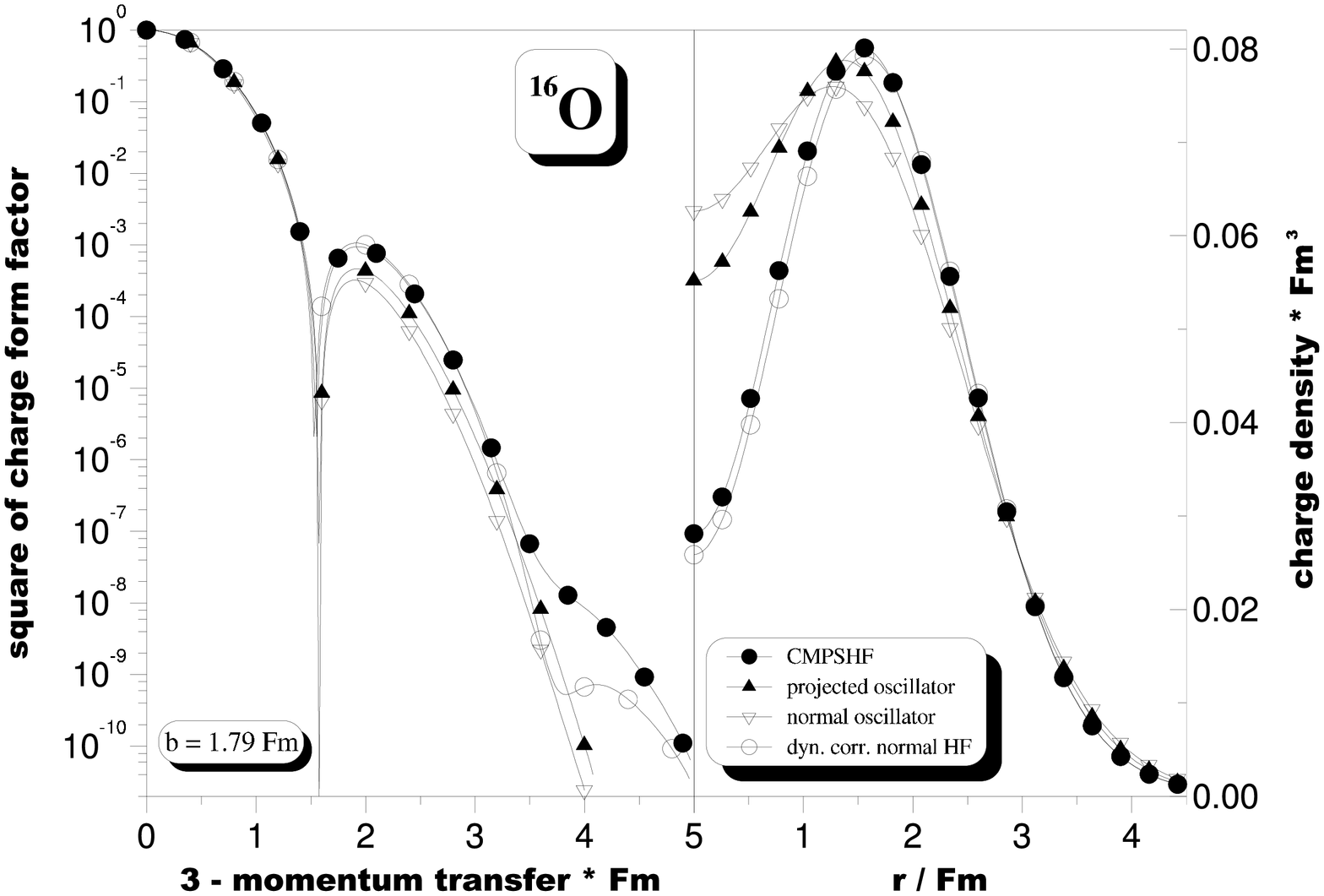}
\caption{Same as in Fig. 7, but for the nucleus $^{16}$O.}
\end{center} 
\end{figure*}

The energy functional (\ref{Eq8}) has to be minimized with respect to
arbitrary variations of the underlying Hartree--Fock transformation $D$.
This transformation, however, has to be unitary and thus not all
of the $M_b\cdot M_b$ matrix elements of $D$ are linear independent.
Nevertheless an unconstrained minimization of the functional (8)
can still be performed, if one parametrizes the underlying Hartree--Fock
transformation $D$ via Thouless' theorem \cite{ref15.}, which states
that any Hartree--Fock determinant $\vert D_d\rangle$ can be represented in
terms of the creation and annihilation operators of some reference
determinant $\vert D_0\rangle$ via
\begin{equation} \label{Eq15}
\vert D_d\rangle\;=\;c(d)\exp\left\{\sum_{p,h}\; d_{ph}
b^{\dag}_p(D_0) b_h(D_0)\right\}\vert D_0\rangle 
\end{equation} 
provided that the two determinants are non-orthogonal, since
\begin{equation} \label{Eq16}
c(d)\;=\;\langle D_0\vert D_d\rangle 
\end{equation}
The creation operators belonging to the
Hartree--Fock determinant $\vert D_d\rangle$ are then related to those of the
reference determinant $\vert D_0\rangle$ via
\begin{equation} \label{Eq17}
b^{\dag}_h(D_d)\;=\;\sum_{h'} [L^{-1}]_{hh'}\left(
b^{\dag}_{h'}(D_0)\;+\;\sum_{p'}\;d_{p'h'} b^{\dag}_{p'}(D_0)
\right)
\end{equation}
for the occupied and
\begin{equation} \label{Eq18}
b^{\dag}_p(D_d)\;=\;\sum_{p'} [M^{-1}]_{pp'}\left(
b^{\dag}_{p'}(D_0)\;-\;\sum_{h'}\;d^*_{p'h'} b^{\dag}_{h'}(D_0)
\right)
\end{equation}
for the unoccupied states, respectively. They are given
in terms of the $(M_b-A)\cdot A$ linear independent variables $d_{ph}$.
The $(A\times A)$ matrix $L$ in (\ref{Eq17}) is defined by the expression
\begin{equation} \label{Eq19}
{\bf 1}_A\;+\;d^T d^*\;=\;LL^{\dag} 
\end{equation}
while the $((M_b-A)\times (M_b-A))$ matrix $M$ out of (\ref{Eq18}) can be
obtained by the solution of the equation
\begin{equation} \label{Eq20}
{\bf 1}_{M_b-A}\;+\;d^* d^T\;=\;MM^{\dag} 
\end{equation}
The variational equations resulting from the minimization of the
functional (\ref{Eq8}) thus get finally the form
\begin{equation} \label{Eq21}
{{\partial E_{pr}}\over{\partial d_{ph}}}\;=\;
\left[{M^{-1}}^{\dag}\; G\; L^{-1}\right]_{ph}\;\equiv\;0 
\end{equation}
where the $((M_b-A)\times A)$ matrix $G$ is defined as
\begin{equation} \label{Eq22}
G_{ph}\;\equiv\;{{\int d a\;a^2\int d\Omega_a\;
g_{ph}(\vec a\,)}\over{
\int d a\;a^2\int d\Omega_a\;
n(\vec a\,)}}
\end{equation}
and the function $g_{ph}(\vec a\,)$ is given by

\begin{eqnarray} \label{Eq23}
g_{ph}(\vec a\,)=
\langle D\vert\;[\hat H\;-\;E_{pr}]\;\hat S(\vec a\,)
b^{\dag}_p(D) b_h(D)\;\vert D\rangle =
\nonumber\\
\sum_{i,r=1}^{M_b}\;\sum_{h'=1}^A\;[X^{-1}(\vec a\,)]_{hh'}
D_{ih'}
\Biggl\{
[\;h(\vec a\,)\;-\;E_{pr} n(\vec a\,)]\;\delta_{ir}+
\nonumber\\
n(\vec a\,)\sum_{k=1}^{M_b}\;\tilde\Gamma_{ik}(\vec a\,)
({\bf 1}\;-\;\tilde\rho(\vec a\,))_{kr}
\Biggr\}
\sum_{s=1}^{M_b}\;S_{rs}(\vec a\,)\;D^*_{sp} 
\nonumber\\
\end{eqnarray}
where 

\begin{equation} \label{Eq24}
\tilde\Gamma_{ik}(\vec a\,)\;\equiv\;
t(ik)\;+\;\sum_{rs}\;v(irks)\;\tilde\rho_{sr}(\vec a\,)
\end{equation}
The ``local'' gradient vector (\ref{Eq22}), obviously,
has to vanish at the solution of (\ref{Eq21}), too. This solution
can be obtained using standard methods
as they have been described, e.g., in Ref. \cite{ref16.}.

\begin{figure*}
\begin{center}
\includegraphics[angle=-90,width=12cm]{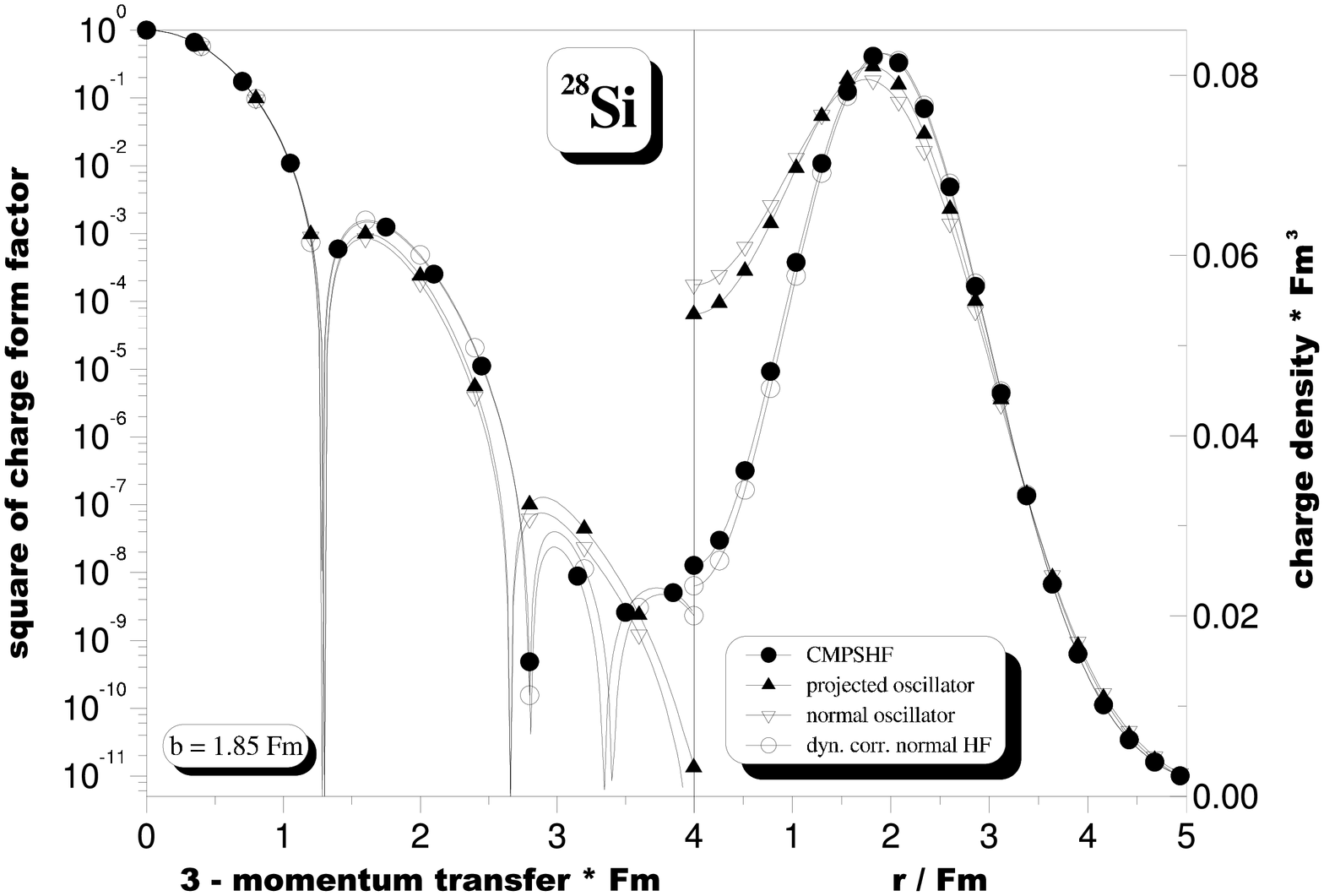}
\caption{Same as in Fig. 7, but for the nucleus $^{28}$Si.}
\end{center} 
\end{figure*}

Up to now no symmetry restrictions have been imposed on the
Hartree--Fock transformation (\ref{Eq3}), (\ref{Eq4}). Thus the Hartree--Fock 
vacuum (\ref{Eq2}) breaks
in general besides the translational invariance also other symmetries like,
e.g., the conservation of the total angular momentum and the parity.
For this general case therefore besides the momentum projection
also the projection on these other symmetries would be required.
The situation becomes, however, much simpler, if only spherically symmetric
Hartree--Fock transformations are admitted. Then each of the selfconsistent
states created by the operators (\ref{Eq3}) has the isospin 3-projection,
the orbital and total angular momentum and the 3-projection of
the latter as ``good'' quantum numbers, and the sums in (\ref{Eq3}) 
and (\ref{Eq4})
run only over the node quantum number. For nuclei with closed
angular momentum subshells the corresponding Hartree--Fock vacuum (\ref{Eq2})
has then total angular momentum $I^\pi = 0^+$ and conserves
the proton as well as the neutron number. Consequently the
projection on these symmetries becomes redundant and we are left
with only the linear momentum projection as described above.

Furthermore, for spherically symmetric systems obviously neither
the shifted overlap (\ref{Eq9}) nor the corresponding energy function
(\ref{Eq11}) do depend on the direction of the shift vector $\vec a$.
Thus the angle integrations in (\ref{Eq8}) and (\ref{Eq21}) induced by the
 operator (\ref{Eq6})
become trivial and only a single integral over the radial
component of the shift vector remains to be done numerically.
An explicit formulation of this special case will not be
given in the present paper. However, it is obvious that the
calculation of the expressions needed for the minimization of the
energy functional (\ref{Eq8}) is then simplified considerably.

\section{Results and discussion.}
We have considered the six nuclei $^4$He, $^{12}$C, $^{16}$O, $^{28}$Si,
$^{32}$S and $^{40}$Ca. As hamiltonian, as in ref. \cite{ref12.}, the 
Brink--Boeker
interaction B1 \cite{ref14.} complemented with a short range (0.5 Fm) two--body
spin--orbit term having the same volume--integral as the corresponding
zero--range term of the Gogny--force D1S \cite{ref17.}, plus the Coulomb force and the
kinetic energy has been used. First, the energy of the simple oscillator
determinants for these nuclei (e.g., $(0s)^4(0p)^{12}$ for $^{16}$O) has been
minimized with respect to the oscillator length--parameter $b$. For the
intermediate states needed to compute the shifted energy function (\ref{Eq11}) here
four major shells more than in the basis have been taken (e.g., in $^{16}$O
the maximum $N=2n+l$ of the oscillator determinant is 1. Hence, for the
intermediate states all orbits up to N=5 have been used). The results
obtained were identical to those obtained analytically in ref. \cite{ref12.}, which 
is a good check of the convergence of the numerical procedure.

In the next step then, for increasing size of the single particle basis
up to $N=2n+l=18$, in each nucleus and each basis system always three
different Hartree--Fock calculations have been performed :

First, a usual spherical Hartree--Fock calculation was done, in which
the energy
\begin{equation} \label{Eq25}
E^{\prime}_n\;=\;\langle D_n\vert\hat H\vert D_n\rangle
\end{equation}
is minimized and after convergence corrected by subtracting the expectation
value of the kinetic energy of center of mass motion
\begin{equation} \label{Eq26}
E_n\;=\;E^{\prime}_n\;-\;\langle D_n\vert{{{\hat P\,}^2}\over{2MA}}
\vert D_n\rangle
\end{equation}
This is the normal approach as indicated by the subscripts $n$ at the
total energy and the wave function.

Second, a corrected spherical Hartree--Fock calculation has been done, in
which the expectation value of the internal hamiltonian
\begin{equation} \label{Eq27}
E_c\;=\;\langle D_c\vert\left(\hat H\,-\,{{{\hat P\,}^2}\over{2MA}}\right)
\vert D_c\rangle
\end{equation}
is minimized. The subscript $c$ refers to this corrected approach.

\begin{figure*}
\begin{center}
\includegraphics[angle=-90,width=12cm]{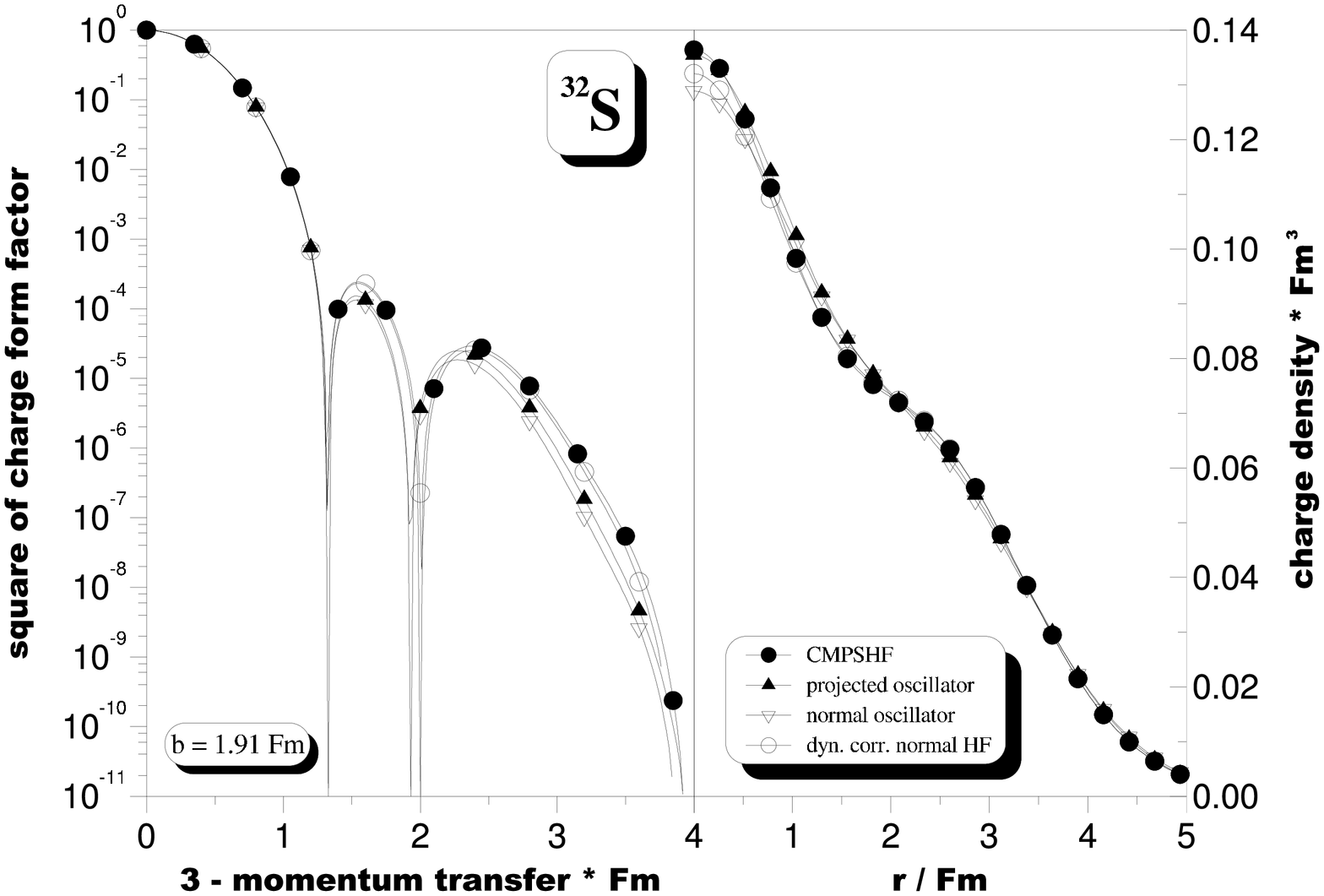}
\caption{Same as in Fig. 7, but for the nucleus $^{32}$S.}
\end{center} 
\end{figure*}

Third, a spherical Hartree--Fock calculation with projection into the center
of mass rest frame before the variation as described in section 2 has been
performed. In this case the energy--functional (\ref{Eq8})
\begin{equation} \label{Eq28}
E_{pr}\;=\;{{\langle D_{pr}\vert\hat H\,\hat C (0)\vert D_{pr}\rangle}
\over{\langle D_{pr}\vert\hat C (0)\vert D_{pr}\rangle}}
\end{equation}
has been minimized. Here, for the intermediate states again always four
major shells more than for the basis have been taken into account.
Note, that $\vert D_n\rangle$, $\vert D_c\rangle$ and $\vert D_{pr}\rangle$
result from different variational calculations and are hence different.

Finally, for the largest basis system ($N=18$), we have studied 
in each nucleus a widely used approximate description to deal with the
center of mass motion : instead of minimizing $E_c$ out of equation (\ref{Eq27})
one minimizes
\begin{eqnarray} \label{Eq29}
E^{\prime}_{\lambda}\;=\;\langle D_{\lambda}\vert\left(\hat H\,-\,
{{{\hat P\,}^2}\over{2MA}}\right)\vert D_{\lambda}\rangle\,+
\nonumber\\
+ \,\lambda\,\cdot\,
\langle D_{\lambda}\vert\left({{{\hat P\,}^2}\over{2MA}}
\,+\,{1\over 2} MA\omega^2{\vec R\,}^2\right)\vert D_{\lambda}
\rangle
\end{eqnarray}
with a large Lagrangian multiplier $\lambda$, i.e., one penalizes center of
mass excitations. This prescription is exact for so called complete
$n\hbar\omega$ configuration spaces \cite{ref2.}, however, is often applied also
in truncated shell--model spaces [18,19]. The internal energy (i.e., (\ref{Eq29})
without the penalizing term) has been obtained for the three different
$\lambda$ values 10, 100 and 10000.

In addition, again always for the largest basis, the hole--energies
obtained for the corrected approach
\begin{equation} \label{Eq30}
E^h_c\;=\;E_c\,-\,\langle D_c\vert b_h^{\dagger}(D_c)\left(
\hat H\,-\,{{{\hat P\,}^2}\over{2M(A-1)}}\right) b_h(D_c)
\vert D_c\rangle
\end{equation}
have been compared with the projected results
\begin{equation} \label{Eq31}
E^h_{pr}\;=\;E_{pr}\,-\,{{\langle D_{pr}\vert b_h^{\dagger}(D_{pr})
\hat H\hat C(0) b_h(D_{pr})\vert D_{pr}\rangle}\over
{\langle D_{pr}\vert b_h^{\dagger}(D_{pr})\hat C(0)
b_h(D_{pr})\vert D_{pr}\rangle}}
\end{equation} 
as well as the corresponding results for the simple oscillator occupations.
For the nuclei, in which two s--states are occupied ($^{32}$S and $^{40}$Ca),
obviously an additional diagonalization has been performed.

Note, that the definition (\ref{Eq30}) differs from the usual expression since via the
kinetic energy of the center of mass motion the internal hamiltonian becomes
A--dependent. The resulting difference with respect to the usual
expression is for non--spurious oscillator hole--states $3\hbar\omega/4(A-1)$
and in the general case always larger than this lower limit. Inserting
the results for $\hbar\omega$ into this formula, one obtains considerable
effects even for the larger A-values considered here.

The results for the total binding energies and the hole-energies of the
considered nuclei are summerized in figures 1 to 6. The left side of each
figure presents the total binding energy as function of the size of the
basis. Three different curves are plotted : open circles refer to the
Hartree--Fock results (\ref{Eq26}) where the kinetic energy of the center of mass
motion is subtracted after the variation, full triangles give the results
of the corrected approach (\ref{Eq27}), in which this subtraction is done before
the variation and full circles display the results of the spherical
Hartree--Fock calculations with projection into center of mass rest frame
before the variation (\ref{Eq28}).

For pure oscillator occupations (i.e., the smallest basis) these three
curves obviously coincide, for larger basis systems, however, they differ
considerably, i.e., display rather different major--shell--mixing. Let us
first concentrate on the unprojected approaches (\ref{Eq26}) and (\ref{Eq27}). As
 expected, the corrected approach (\ref{Eq27}) yields always a lower binding energy than the
normal one (\ref{Eq26}), however, for all but one of the considered nuclei
the corresponding curves run almost parallel with increasing basis size 
and their difference is rather small. The exception is the case of $^4$He,
where (\ref{Eq26}) even yields a decrease in binding energy with the basis size
such indicating that the underlying wave functions have a rather different
structure than those obtained via (\ref{Eq27}). The energy gain of the projected
approach (\ref{Eq28}) with respect to the corrected prescription (\ref{Eq27}) is in all
considered nuclei (except $^4$He) much larger than that of the latter 
with respect to (\ref{Eq26}). For $^{40}$Ca in the largest basis system, e.g., the
projected binding energy is 1.25 MeV lower than the corrected result, while
the latter is only 136 keV lower than the normal one. Note, that these 1.25
MeV amount to almost 20 percent of the total major shell mixing obtained
in the corrected approach (\ref{Eq27}). Thus, obviously, the restoration of Galilean
invariance yields a considerable effect on the total binding energy
and should not be neglected even for nuclei as heavy as $^{40}$Ca.

Furthermore, as can be seen from the inverted triangles in the figures, 
the prescription (\ref{Eq29}), which penalizes center of mass excitations, fails
completely. For the largest Lagrange multiplier ($\lambda=10000$) 
the procedure yields in all considered nuclei just the simple oscillator
occupation. This, definitely, is a non--spurious state (i.e., contains
no center of mass excitations), however, the major--shell--mixing is
completely supressed in this solution. This is a severe warning to use
the prescription (\ref{Eq29}) in incomplete model spaces : it always prefers
non--spurious (one valence shell) solutions and is hence uncontrollable
even if the configuration space is less severly truncated as in the
simple Hartree--Fock approach discussed here.

On the right side of figures 1 to 6 we display the proton and neutron 
hole--energies in the considered nuclei. Always the corrected results (\ref{Eq30})
(indicated by the label no) are compared with the projected energies (\ref{Eq31})
(indicated by the label pr) for both the oscillator occupation (HO) as well as
the Hartree-Fock approach (HF) calculated in the $N=18$ basis. 

Though the underlying wave functions (and total binding energies)
are considerably different, in all
considered nuclei the harmonic oscillator approach and the Hartree--Fock
method yield remarkably similar results. For the non--spurious hole--states 
out of the last occupied major shell in the harmonic oscillator approach
the corrected and projected results have to be identical as demonstrated
analytically in ref. \cite{ref12.} and this feature holds to a large extent for
the Hartree--Fock results, too. For the hole--states out of the second
and third but last occupied shell corrected and projected results display
in both approaches rather similar pronounced differences. So, e.g.,
in $^{16}$O the projected p--holes are more than 6 MeV lower in energy
than the corresponding corrected results and even in $^{40}$Ca the differences
are still about 2.5 MeV for both the p-- and the lowest s--holes. It was
demonstrated in ref.\cite{ref12.} that these differences are consistent with the
differences in the spectroscopic factors out of ref.\cite{ref10.}. This will be
discussed in more detail in the second of the present series of papers.

\begin{figure*}
\begin{center}
\includegraphics[angle=-90,width=12cm]{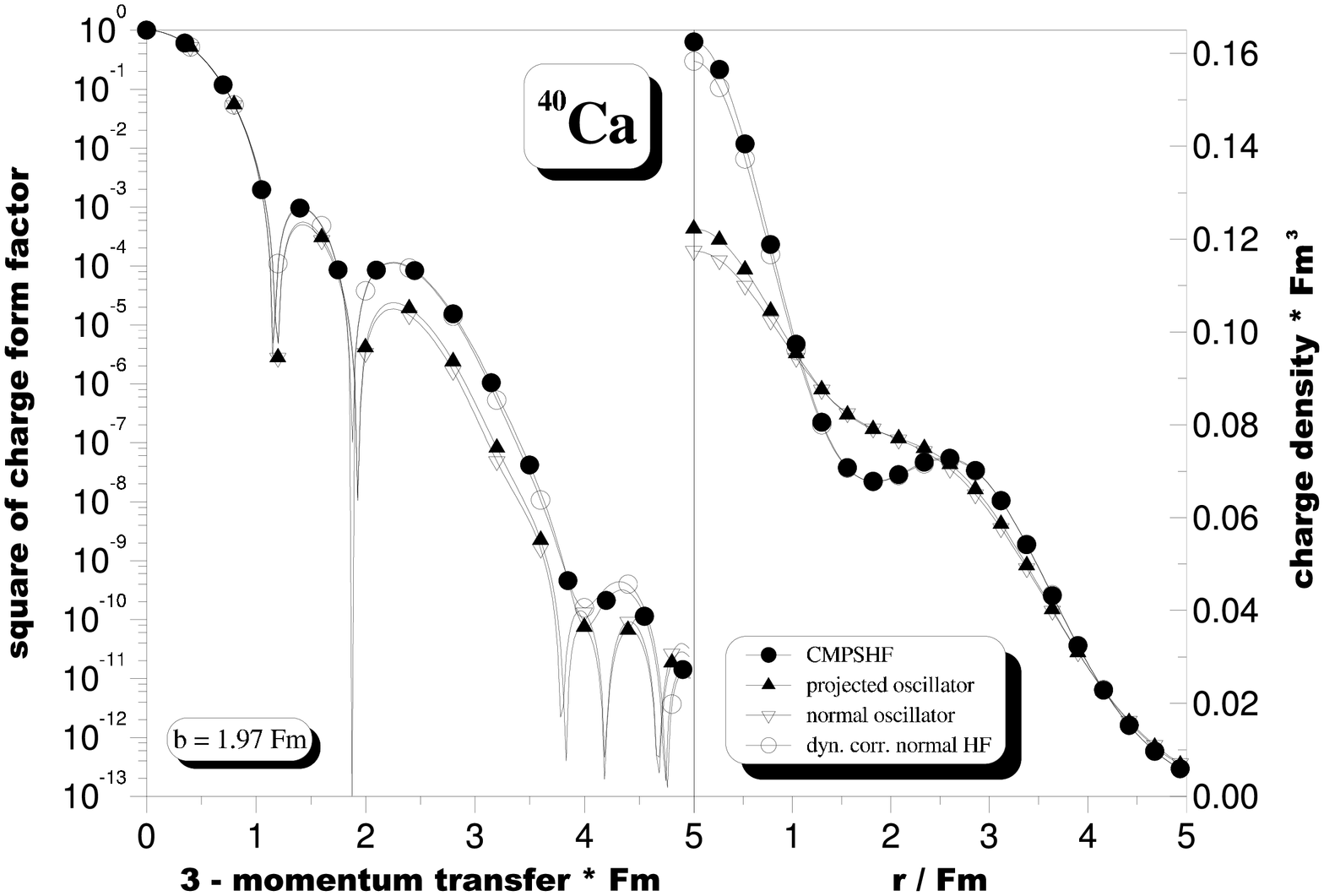}
\caption{Same as in Fig. 7, but for the nucleus $^{40}$Ca.}
\end{center} 
\end{figure*}

An interesting observation is made for the two nuclei $^{28}$Si and
$^{32}$S. Here, the p1/2--holes are almost unaffected by the projection,
while the p3/2--holes show the same differences as, e.g., observed in
$^{40}$Ca. Since the coupling of p1/2 and d5/2 to angular momentum one is
not possible and the d3/2--orbit is unoccupied in these two nuclei,
this observation points to the dominance of angular momentum one
couplings for the hole energies.

The figures 7 to 12 demonstrate the effects on the charge form factors and
corresponding charge densities. Usually, the operator for the charge
density in momentum representation is written as \cite{ref20.,ref11.}
\begin{equation} \label{Eq32}
\hat\rho_n\,\equiv\,\sum_{\tau} f_{\tau}(Q^2)\,\sum_{i=1}^{N_{\tau}}\,
\exp\{i\vec q\cdot\vec r_i\} 
\end{equation}
where $\tau$ is the isospin projection (proton or neutron) and
the nucleon charge form factors $f_{\tau}$ are given by
\begin{equation} \label{Eq33}
f_{\tau}(Q^2)\,\equiv\,G^{\tau}_E(Q^2)\,-\,{{Q^2}\over{8M^2}}\,
{{G^{\tau}_E(Q^2)\,+\,{{Q^2}\over{4M^2}}\,G^{\tau}_M(Q^2)}\over
{1\,+\,{{Q^2}\over{4M^2}}}}
\end{equation}
with the Sachs--form factors parametrized in the well--known
dipole form (see, e.g., \cite{ref21.})

\begin{eqnarray} \label{Eq34}
G^p_E(Q^2)\,\equiv\,\left[1\,+\,{{Q^2}\over{(843 {\rm MeV})^2}}
\right]^{-2}
\nonumber\\
G^{\tau}_M(Q^2)\,\equiv\,\mu_{\tau}\,G^p_E(Q^2)
\nonumber\\
\mu_p\,=+2.793
\nonumber\\
\mu_n\,=-1.913
\nonumber\\
G^n_E(Q^2)\,\equiv\,-\mu_n\,{{Q^2}\over{4M^2}}\,{1\over{1\,+\,5.6
{{Q^2}\over{4M^2}}}}\,G^p_E(Q^2)
\end{eqnarray}
Here $M$ is the nucleon mass and $Q^2$ the negative square of the
4--momentum transfer
\begin{equation} \label{Eq35}
Q^2\,\equiv\,(\hbar c \vec q\,)^2\,-\,(\Delta E)^2
\noindent
\end{equation}
with $\Delta E$ being the energy transfer and $\vec q$ the 3--momentum
transfer to the system. For elastic electron scattering the energy
transfer is given by the recoil energy $(\hbar c\vec q\,)^2/(2AMc^2)$
so that here
\begin{equation} \label{Eq36}
Q^2\,=\,(\hbar c\vec q\,)^2\left\{1\,-\,{{(\hbar c\vec q\,)^2}\over
{4A^2 M^2 c^4}}\right\}
\end{equation}
If, as in our case, the ground state is described by a single determinant
$\vert D\rangle$, then the ``normal'' elastic charge form factor has the form
\begin{equation} \label{Eq37}
F_{ch}^n(Q^2)\,=\,\langle D\vert\hat\rho_n\vert D\rangle
\end{equation}
and the corresponding charge density is just the Fourier--transform of
this expression.

Obviously, to obtain a translational invariant form for the charge density
operator, (\ref{Eq32}) has to be complemented with the so--called
Gartenhaus--Schwartz operator as has been demonstrated in ref. \cite{ref11.}
\begin{equation} \label{Eq38}
\hat\rho_{inv}\,\equiv\,\hat\rho_n\,\exp\{-i\vec q\cdot\vec R\}
\end{equation}
The Galilei--invariant form of the charge form factor is thus
\begin{equation} \label{Eq39}
F_{ch}^{pr}(Q^2)\,=\,{{\langle D\vert\hat\rho_{inv}\hat C(0)\vert
D\rangle}\over{\langle D\vert\hat C(0)\vert D\rangle}}
\noindent
\end{equation}
The matrix elements needed to compute this expression are given in
appendix B. The corresponding charge density is then again obtained by
Fourier--transforming this expression.

It has been demonstrated already some time ago \cite{ref8.} that using the
so--called Gaussian--overlap--approximation for both the shift as well as
for the Gartenhaus--Schwartz operator (\ref{Eq38}) reduces to the
``dynamically corrected'' charge form factor
\begin{equation} \label{Eq40}
F_{ch}^{dy}(Q^2)\,=\,F_{ch}^n(Q^2)\,\exp\left\{{3\over 8}\,{{{\vec q\,}^2}
\over{\langle D\vert{\hat P\,}^2\vert D\rangle}}\right\}
\noindent
\end{equation}
In case that $\vert D\rangle$ is a non--spurious oscillator state the
exponential factor in (40) gets the form $\exp\{(\vec q\,b/2)^2/A\}$,
which is the famous Tassie--Barker correction [5].

On the left side of the figures 7 to 12 we compare for the considered
nuclei the normal form factor (\ref{Eq37}) for the oscillator occupation (inverted
open triangles) with the corresponding projected one (\ref{Eq39}) (full triangles),
the dynamically corrected one (\ref{Eq40}), resulting from the solution
$\vert D_c\rangle$ of the minimization of the Hartree--Fock energy
functional (\ref{Eq27}), and, finally, the Galilei--invariant one (\ref{Eq38}) computed
from the solution $\vert D_{pr}\rangle$ of the minimization of the
projected energy functional (\ref{Eq28}) (full circles). The corresponding
charge densities are given on the right side of the figures. All
Hartree--Fock results have been obtained using the largest basis with
19 major oscillator shells.

Let us first concentrate on the oscillator occupation. In $^4$He
a large difference between the normal and the projected oscillator
form factors at high momentum transfer and consequently for the
charge density at small radii is observed. Since we have here a 
non--spurious oscillator state this difference is entirely due
to the Tassie--Barker correction. This correction decreases with
increasing mass number and can in $^{40}$Ca almost be neglected.
On the other hand the difference of the Hartree--Fock results
with respect to the oscillator ones increase with increasing mass number
due to the increasing major shell mixing. E.g., in $^{40}$Ca Hartree--Fock
and oscillator results look rather different.

\begin{figure*}
\begin{center}
\includegraphics[angle=-90,width=12cm]{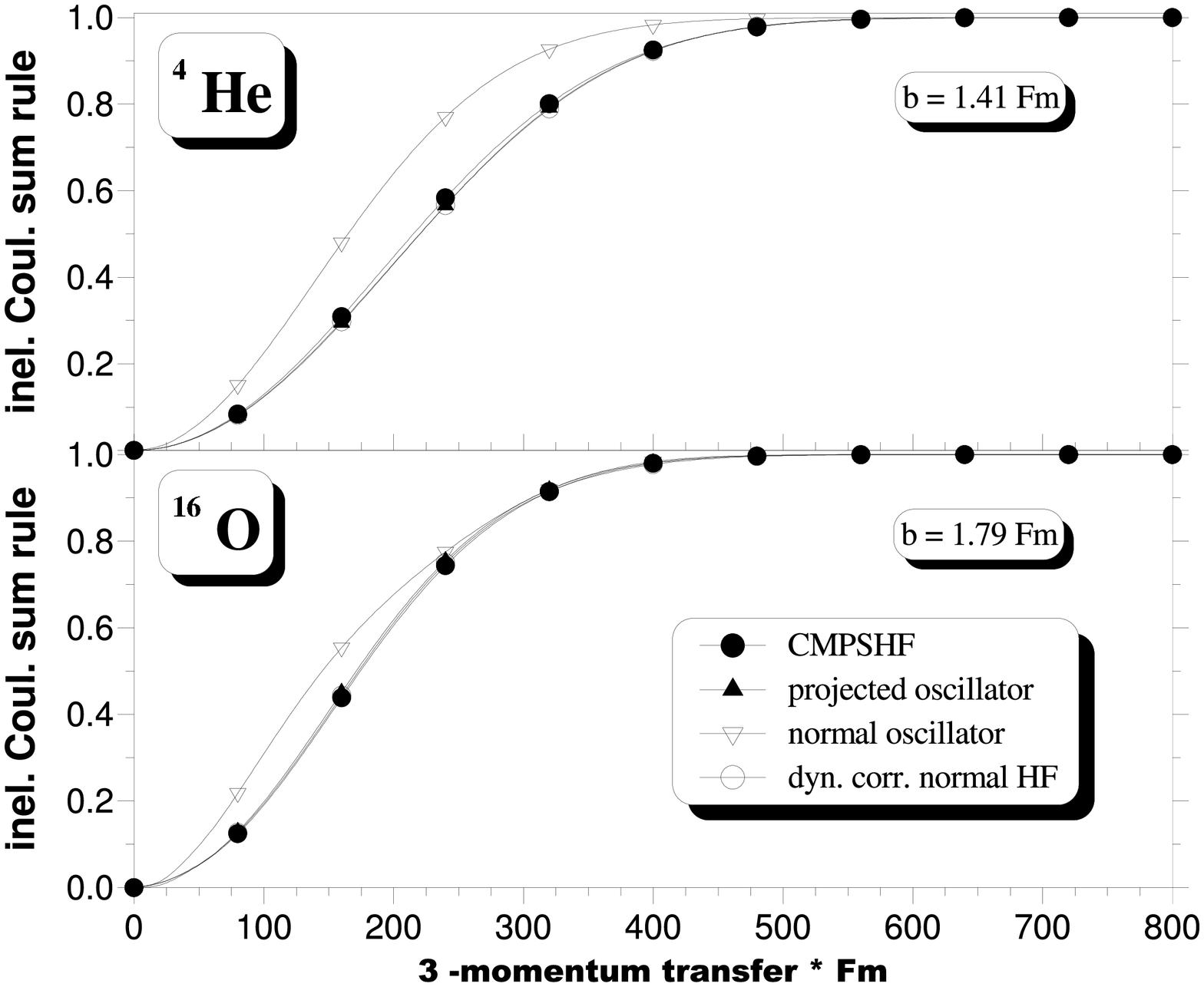}
\caption{The inelastic (mathematical) Coulomb sum rule as function
of the 3--momentum transfer. Compared are the results of calculations
using an oscillator occupation and an uncorrected elastic charge
form factor (for point protons) with those using the same occupation
but including the Tassie--Barker factor for the elastic charge form factor,
those of a normal Hartree--Fock calculation including the dynamical
correction for the elastic charge form factor and, finally, those
of the projected CMPSHF approach. The upper part of the figure gives the results for
$^4$He, the lower part those for the nucleus $^{16}$O.}
\end{center} 
\end{figure*}

Though computed with rather different wave functions the projected
and dynamically corrected form factors and charge densities display
only rather small differences in all the considered nuclei except
$^4$He. This is somewhat surprising since in ref. \cite{ref7.,ref8.} 
larger effects
of the projection have been seen even for nuclei up to A=40. However,
besides being limited to projection after the variation, these
calculations had been done with different effective interactions
than used in the present work. Thus the present results do not
indicate that for the elastic form factors and charge densities the
dynamical correction is good enough and no projection is needed.
Instead a more careful study using various effective interactions
is required.

In addition to the form factors we have studied the mathematical Coulomb
sum rules, too. Details of its definition can be found in ref. \cite{ref11.}.
As usual we assumed point nucleons, i.e., we set the
nucleon form factors out of eq. (\ref{Eq33}) to 1 for the proton and 0 for the
neutron. Furthermore, again as usual, we subtracted the square of the
elastic form factor and divided the result by the charge number in
order to obtain the so--called inelastic Coulomb sum rules. Without
any center of mass correction these have the form
\begin{equation} \label{Eq41}
\Sigma^{inel,\,n}(q)\,=\,{1\over Z}\,\left\{
\langle D_n\vert\hat\rho_n{\hat\rho_n\,}^{\dag}\vert D_n\rangle\,-\, 
\left(F_{ch}^n(q)\right)^2\right\} 
\end{equation}
where the elastic form factor out of eq. (\ref{Eq37}) has to be taken in the
point proton limit. If we include the dynamical correction for the
elastic form factor we obtain the result
\begin{equation} \label{Eq42}
\Sigma^{inel,\,dy}(q)\,=\,{1\over Z}\,\left\{
\langle D_c\vert\hat\rho_n{\hat\rho_n\,}^{\dag}\vert D_c\rangle\,-\, 
\left(F_{ch}^{dy}(q)\right)^2\right\}
\noindent
\end{equation}
with the point proton limit for expression (\ref{Eq40}),
and, finally, the linear momentum projected expression has the form
\begin{equation} \label{Eq43}
\Sigma^{inel,\,pr}(q)\,=\,{1\over Z}\,\left\{
{{\langle D_{pr}\vert\hat\rho_n{\hat\rho_n\,}^{\dag}\hat C(0)\vert D_{pr}
\rangle}
\over{\langle D_{pr}\vert\hat C(0)\vert D_{pr}\rangle}}\,-\, 
\left(F_{ch}^{pr}(q)\right)^2\right\}
\end{equation}
with the point proton limit of the form factor (39). Explicit forms for the
matrix elements entering the expressions (41-43) are for spherically symmetric
determinants $\vert D\rangle$ given in appendix C.
Note, that it is irrelevant whether in the first term of
these expression the normal (32) or the invariant
form (38) of the charge density operator is used, since in these matrix
elements the Gartenhaus--Schwartz operator does drop out. As for the charge
form factors and densities the normal (41) and projected (43) results for
the oscillator occupation have been compared with the dynamically
corrected (42) and the projected (43) Hartree--Fock results. For the
computation of the latter obviously again the solutions $\vert D_c\rangle$
and $\vert D_{pr}\rangle$ of the corresponding variational calculations
have been taken.

The results for $^4$He and $^{16}$O are displayed in figure 13. Plotted
are the inelastic sum rules as defined above as functions of the
3--momentum transfer $q$. As expected from the similarity of the
dynamically corrected and the projected elastic Hartree--Fock
form factors almost no differences between these two approaches are obtained
for the inelastic sum rules either. That the projected oscillator
results almost coincide with the Hartree--Fock results, too, is a clear
indication that the inelastic sum rule is rather insensitive to the
major shell mixing. However, all these results approach the limit of 1
considerably slower than the normal oscillator approach. This difference
is entirely due to the square of the elastic form factor in the above
expressions and demonstrates that a correct treatment of the latter
(either exact by projection or approximate by the dynamical correction)
is definitely required.

Since for the inelastic sum rules of the other considered nuclei
the same behaviour as demonstrated in figure 13 is obtained, we shall
not discuss them here.

\section{Conclusions.}
We have presented the total binding energies, hole energies, form factors and
charge densities as they result from spherical Hartree--Fock calculations
with projection into the center of mass rest frame before the variation
for the six nuclei $^4$He, $^{12}$C, $^{16}$O, $^{28}$Si, $^{32}$S and
$^{40}$Ca and have compared them to the standard Hartree--Fock results
obtained by subtracting the kinetic energy of the center of mass motion
either after or before the variation. Furthermore, for the two nuclei
$^4$He and $^{16}$O, we have discussed the inelastic Coulomb sum rule
resulting from these different approaches. For comparison, in addition
the results for pure oscillator occupations have been discussed.

For the total binding energies considerable effects of the correct
treatment of Galilei--invariance are seen. In all the considered nuclei
the energy gains of the momentum projected solutions with respect to the
conventionally corrected approach using just the internal hamiltonian 
(which contains already the usual 1/A effect) in the variation
are a considerable portion of the gains due to major shell mixing and
hence as important as the latter. It was furthermore demonstrated that
the often used approximate prescription to penalize center of mass
excitations by an additional term in the variation does not work at
all at least in the severly truncated configuration spaces used here.
There are strong indications that this prescription does only work
in complete $n\hbar\omega$--spaces and is uncontrollable even if used
in less severly truncated shell--model spaces.

For the hole energies essentially the same features as in ref. \cite{ref12.}
are observed. While the energies of the holes out of the last occupied
shell are almost unaffected, the projected energies of the holes out of
the second and third but last shell are considerably different from their
conventionally corrected counterparts (which again include the trivial 1/A
effects).

For the elastic charge form factors and densities (except for the lightest
considered system) there are little differences obtained between the
projected and the dynamically corrected approach though these two
approaches use rather different wave functions resulting from
different variational calculations and the same holds for the inelastic
Coulomb sum rules. However, these results may be changed, if a different
(more realistic) effective interaction is used, and hence have to be 
interpreted with some care.

In conclusion, it has been demonstrated that a correct treatment of
Galilei--invariance in the nuclear many--body problem is possible
via projection methods and that its effects are not only important
for simple harmonic oscillator configurations as shown in refs. \cite{ref10.,ref11.,ref12.,ref13.}
but also for more realistic wave functions. We shall show in the second
of these two papers that this holds also for the spectral functions
and spectroscopic factors. Thus we think that the (up to now mostly
neglected) restoration of Galilei--invariance is unavoidable in future
nuclear structure calculations, and, on the long run, should also be
done in more sophisticated approaches like the shell--model [18],
the quantum Monte--Carlo diagonalization method \cite{ref19.} or the VAMPIR 
approach \cite{ref22.}.

\begin{acknowledgement}
We are grateful that the present study has been supported by the Deutsche 
Forschungsgemeinschaft via the contracts FA26/1 and FA26/2.
\end{acknowledgement}

\section{Appendix A: Oscillator matrix elements of the shift operator.}
The single particle matrix elements $S_{ik}(\vec a\,)$ (10) of the
shift operator (\ref{Eq7}) within oscillator single particle states
play an essential role in the projected formalism presented
in section 2. They have been given already in ref. \cite{ref6.}. Using the
usual quantum numbers $\tau$, $n$, $l$, $j$ and $m$ for the isospin
projection, the node number (starting from zero), the orbital angular
momentum $l$, which is coupled with the spin to the total angular
momentum $j$ and its 3--projection $m$, we obtain

\begin{eqnarray} \label{Eq1AP}
\langle\tau_1 n_1 l_1 j_1 m_1\vert\hat S(\vec a\,)\vert
\tau_2 n_2 l_2 j_2 m_2\rangle 
=
\delta_{\tau_1 \tau_2}
\exp\left\{-{1\over 4}\alpha^2\right\} 
\nonumber\\
\times \sum_L\sqrt{{{4\pi}\over{2L+1}}}\,Y^*_{L\Lambda}(\Omega_a)\,
{1\over 2}\left[1\,+\,(-)^{l_1+l_2+L}\right]
\nonumber\\
\times (-)^{[L+l_2-l_1]/2} 
 \sqrt{(2j_1+1)(2j_2+1)}(-)^{j_1-1/2}
\nonumber\\
\times (j_1 j_2 L | 1/2 -1/2  0) (-)^{j_2-m_2}
\nonumber\\
\times (j_1 j_2 L\vert m_1 -m_2 \Lambda)\,\eta_L^{n_1 l_1 n_2 l_2}
(\alpha)
\nonumber\\
\end{eqnarray}
where $\alpha\,=\,\vert\vec a\,\vert/b$ with $b$ being the oscillator
length and 
\begin{eqnarray} \label{Eq2AP}
\eta_L^{n_1 l_1 n_2 l_2}(\alpha) =
\exp\left\{+{1\over 4}\alpha^2\right\}
(-)^{n_1+n_2}\int\limits_0^{\infty}d\kappa\,{\rm e}^{-\kappa^2} 
\nonumber\\
\times 
\kappa^2\,\tilde R_{n_1 l_1}(\kappa)\,j_L(\kappa\alpha)\,
\tilde R_{n_2 l_2}(\kappa)
\end{eqnarray}
where $\tilde R_{n l}(\kappa)$ are the (dimensionless) polynomial parts of the
usual spherical radial oscillator functions depending on the dimensionless
variable $\kappa$. An analytical form of the expression (\ref{Eq2AP}) has 
been given 
in ref. \cite{ref6.} and will not be repeated here. 

In case that the shift vector can be put in z--direction as in the
spherically symmetric systems considered here, then
\begin{equation} \label{Eq3AP}
\sqrt{{{4\pi}\over{2L+1}}}\,Y^*_{L\Lambda}(\hat z)\,\equiv\,
\delta_{\Lambda 0}
\end{equation}
Then, in eq. (\ref{Eq1AP}), obviously, $m_1$ and $m_2$ have to be equal
and the evaluation of the formulas in section 2 is simplified
considerably.

\section{Appendix B:The projected charge form factor.}
In this appendix we shall give the formulas needed to evaluate the
projected charge form factor out of eq. (39). Again we assume that
the determinant $\vert D\rangle$ is spherical symmetric. This allows
to fix the direction of the momentum transfer to the z--axis.
Furthermore it can be shown easily, that the dependence on the
angle $\varphi_a$ of the shift vector is is trivial and can be
integrated out analytically. Left to be done is then a two--fold
integration over the length of the shift vector and over the angle
$\vartheta_a$ between the shift vector and the z--axis (direction
of the momentum transfer). After some tedious but straightforward
calculation we obtain

\begin{eqnarray}  \label{Eq1AP_B}
F^{pr}_{ch}(Q^2) = {{4\pi b^3}\over{\langle D\vert\hat C(0)
\vert D\rangle}}\exp \left\{-{{A-1}\over{A}}\left({{bq}\over{2}}\right)^2
\right\}
\nonumber\\
\int _{0}^{\infty} d\alpha \alpha^2 \exp \left\{-{{A}\over{4}} \alpha^2\right\}
\int_{0}^{\pi/2} d\vartheta_a \sin\vartheta_a 
\nonumber\\
\cdot\,2{\rm Re}\Biggl\{\Bigl[\prod\limits_{\tau=p,n}
{\rm det} z^{\tau}(bq,\,\alpha,\,\vartheta)\Bigr]\Bigl[\sum\limits_{\tau}
f_{\tau}(Q^2)
\nonumber\\
{\sum\limits_{h_1,h_2 > 0}}^{(\tau)}\Bigl\{
\qquad\;\;\;\, y^{\tau}_{h_1 h_2}(bq,\,\alpha,\,\vartheta)
{z^{\tau}}^{-1}_{h_2 h_1}(bq,\,\alpha,\,\vartheta) 
\nonumber\\
\qquad -\,y^{\tau}_{h_1 {\bar h}_2}(bq,\,\alpha,\,\vartheta)
{z^{\tau}}^{-1}_{h_2 {\bar h}_1}(bq,\,\alpha,\,\vartheta)\Bigr\}
\Bigr]\Biggr\}
\nonumber\\
\end{eqnarray}
where $b$ is again the oscillator length, $\alpha\,=\,\vert\vec a\,\vert/b$,
and $\bar h$ denotes the time reversed partner of the hole state $h$. The
second sum in eq. (\ref{Eq1AP_B}) is restricted to positive values of the 3--projections
of the two hole states. Furthermore $\vec q$ denotes the 3--momentum
transfer while $Q^2$ is the (negative) square of the 4--momentum transfer
as in section 2. The nucleon form factors $f_{\tau}(Q^2)$ are given by 
eq. (\ref{Eq33}). Furthermore

\begin{eqnarray} \label{Eq2AP_B}
z^{\tau}_{12}(bq,\,\alpha,\,\vartheta) =
\sum\limits_L\sqrt{(2j_1+1)(2j_2+1)}(-)^{j_2-m_2}
\nonumber\\
(j_1 j_2 L \vert m_1 -m_2 \Lambda)
(-)^{j_1-1/2}(j_1 j_2 L\vert 1/2 -1/2 0)
\nonumber\\
\Biggl\{{{\sqrt{2L+1}}\over{(l_1 l_2 L\vert 000)}}{1\over 2}
\left[1+(-)^{l_1+l_2+L}\right]
- {{\sqrt{2L+1}}\over{(l_1 l_2 L\vert 1 -1 0)}} \times 
\nonumber\\
\times {1\over 2}
\left[1-(-)^{l_1+l_2+L}\right]\,{{L(L+1)-\kappa_{12}(\kappa_{12}+1)
}\over{2\sqrt{l_1(l_1+1)
l_2(l_2+1)}}}\Biggr\}
\nonumber\\
\sum\limits_{L_1 L_2 l}(L L_2 L_1\vert\Lambda -\Lambda 0)
d^{L_2}_{\Lambda 0}(\vartheta_a)\,i^{L_1}\,(-)^{[l+l_2-L_2]/2}
\nonumber\\
\sum\limits_n (-)^{n_1+n}
\eta_{L_1}^{n_1 l_1 n l}(qb/A)\eta_{L_2}^{n l n_2 l_2}(\alpha)
\sqrt{{{(2L_1+1)(2L_2+1)}\over{2l_2+1}}}
\nonumber\\
(l_1 L_1 l\vert 000)\Bigl\{
(l_1 L l_2\vert 000)(l_1 L_1 l\vert 000)(L L_1 L_2\vert 000)
\nonumber\\
+2\sum\limits_{\lambda=1}^{{\rm min}(l_1,L,L_1)}(-)^{\lambda}
(l_1 L l_2\vert \lambda -\lambda 0)
\nonumber\\
(l_1 L_1 l\vert 
\lambda -\lambda 0)(L L_1 L_2\vert \lambda -\lambda 0)\Bigr\}
\nonumber\\
\end{eqnarray}
where the $\eta$'s are given by expression (\ref{Eq2AP}) and
\begin{equation}
\kappa_{12}\,\equiv\,(l_1-j_1)(2j_1+1)+(l_2-j_2)(2j_2+1)
\end{equation}
The matrix elements of $y^{\tau}_{12}$ have exactly the same form as 
(\ref{Eq2AP_B})
except that the imaginary unit $i$ has to be replaced by $-i$ and the 
argument in the first $\eta$ has to be multiplied with a factor $(A-1)$. Note,
that the expression (\ref{Eq2AP_B}) includes both natural and 
unnatural parity terms in
the sum over $L$. The latter had been neglected in ref. \cite{ref8.}.

\section{Appendix C:The mathematical Coulomb sum rule.}
In this appendix we give the explicit formulas for the matrix elements
entering expressions (\ref{Eq41}-\ref{Eq43}) for the inelastic Coulomb sum rules. In the
normal approach one obtains for spherically symmetric Hartree--Fock
transformations

\begin{eqnarray} \label{Eq1AP_C}
\Sigma_0^{nor}(q) =
\langle D\vert {\hat\rho}_n{\hat\rho}_n^{\dagger}\vert D\rangle 
=\,Z\,+\,\exp\left\{-{1\over 2}(bq)^2\right\}
\nonumber\\
\Biggl\{\Bigl[
{\sum\limits_{\alpha_h l_h j_h}}^{(p)} (2j_h+1)\chi_0^{\alpha_h l_h;\,
\alpha_h l_h}(qb)\Bigr]^2
\nonumber\\
-\,{\sum\limits_{\alpha_h l_h j_h}}^{(p)}
{\sum\limits_{\alpha_{h^{\prime}} l_{h^{\prime}} j_{h^{\prime}}}}^{(p)}
\sum\limits_L\,{1\over 2}\Bigl[1+(-)^{l_h+l_{h^{\prime}}+L}\Bigr]
\nonumber\\
\Delta(l_h,\,l_{h^{\prime}},\,L)(j_h j_{h^{\prime}} L\vert 1/2 -1/2 0)^2
\nonumber\\
(2j_h+1)(2j_{h^{\prime}}+1)\Bigl(\chi_L^{\alpha_h l_h;\,\alpha_{h^{\prime}}
l_{h^{\prime}}}(qb)\Bigr)^2\Biggr\}
\nonumber\\
\end{eqnarray}
where $\Delta(l_h,\,l_{h^{\prime}},\,L)=1$ if $\vert\, l_h-l_{h^{\prime}}\vert
\leq L\leq l_h+l_{h^{\prime}}$ and $=0$ else,

\begin{eqnarray}\label{Eq2AP_C}
\chi_L^{\alpha_h l_h;\,\alpha_{h^{\prime}} l_{h^{\prime}}}(qb)=
{\sum\limits_n}^{(l_h j_h)}{\sum\limits_{n^{\prime}}}^{(l_{h^{\prime}}
j_{h^{\prime}})}\, D_{n\alpha_h}^{p\,l_h j_h}
\nonumber\\
 (-)^{n+n^{\prime}}
\eta_L^{n l_h n^{\prime} l_{h^{\prime}}}(qb)
D_{n^{\prime}\alpha_{h^{\prime}}}^{p\,l_{h^{\prime}} j_{h^{\prime}}},
\noindent
\end{eqnarray}
and the $\eta$'s are given by the expression (\ref{Eq2AP}).

In order to evaluate the corresponding Galilei--invariant expression for
spherically symmetric determinants $\vert D\rangle$ the shift vector can
again be put in z--direction. We obtain

\begin{eqnarray} \label{Eq3AP_C}
\Sigma_0^{proj}(q) =
{{\langle D\vert {\hat\rho}_n{\hat\rho}_n^{\dagger}\hat C(0)\vert D\rangle}
\over{\langle D\vert\hat C(0)\vert D\rangle}}
=\,Z\,+\,\exp\left\{-{1\over 2}(bq)^2\right\} 
\nonumber\\
{\langle D\vert\hat C(0)\vert D\rangle}^{-1}\,
4\pi b^3\,\int\limits_0^{\infty} d\alpha
\,\alpha^2 \langle D\vert\hat S(\hat e_z\cdot\vec a\,)\vert D\rangle
\nonumber\\
\Biggl\{\sum\limits_L {1\over{2L+1}}
\Biggl({\sum\limits_{AC}}^{(p)} M_{AC}^L(qb)\,\tilde\rho_{CA}^L(\alpha)
\Biggr)^2
\nonumber\\
-\,{\sum\limits_{ABCD}}^{(p)}\sum\limits_L\Biggl(
\sum\limits_I (-)^{I-L+1}\left\{\matrix{ j_A &j_C &L\cr j_B &j_D &I\cr}\right\}
\nonumber\\
M_{AD}^I(qb)M_{BC}^I(qb)\Biggr){\tilde\rho}_{CA}^L(\alpha)
{\tilde\rho}_{DB}^L(\alpha)\Biggr\}
\nonumber\\
\end{eqnarray}
where $A,\,B$,... denote the quantum numbers of the oscillator single particle
basis states $n_A l_A j_A,\,n_B l_B j_B$,..., $\vec\alpha\,=\,
\vec a/b$ with $b$ being the oscillator length parameter, and

\begin{eqnarray} \label{Eq4AP_C}
M_{AC}^L(qb)={1\over 2}\left[1+(-)^{l_A+l_C+ L}\right]
\Delta(l_A,\,l_C,\,L)
\nonumber\\
\sqrt{(2j_A+1)(2j_C+1)}
(-)^{j_A-1/2}(j_A j_C L\vert 1/2 -1/2 0)
\nonumber\\
(-)^{n_A+n_C}\,\eta_L^{n_A l_A n_C l_C}(qb)
\nonumber\\
\end{eqnarray}
are the reduced oscillator single particle matrix elements of the normal
charge density operator in momentum representation. The supersript $(p)$
at the sum symbols means that only proton--orbits are considered. Finally,
\begin{equation} \label{Eq5AP_C}
{\tilde\rho}_{CA}^L(\alpha)\,\equiv\,\sum\limits_m
(-)^{j_C-m}(j_A j_C L\vert m -m 0)\,{\tilde\rho}_{CA}^{(m)}(\alpha)
\end{equation}
where
\begin{eqnarray} \label{Eq6AP_C}
{\tilde\rho}_{CA}^{(m)}(\alpha)=
{\sum\limits_{HH^{\prime}}}^{(p)}\, S_{CH}^{(pm)}(\alpha)
{S_{HH^{\prime}}^{(pm)}}^{-1}(\alpha)  
\nonumber\\
D_{n_A\alpha_{H^{\prime}}}^{(p\,l_A
j_A)}\,\delta_{l_{H^{\prime}} l_A}\delta_{j_{H^{\prime}} j_A}
\end{eqnarray}
In eq. ( \ref{Eq6AP_C}) use has been made of the fact, that the single particle matrix
elements do not mix different isospin projections, and, for the shift
vector in z--direction, do not mix states with different total angular
momentum projections $m$ either. This is indicated by the superscipts
$(pm)$.

\end{document}